\documentclass[aps,prc,a4paper,nofootinbib,showpacs,showkeys,nofootinbib,preprintnumbers,superscriptaddress]{revtex4}

\usepackage{amsmath}
\usepackage{amssymb}
\usepackage{bm}
\usepackage{graphicx}
\usepackage{color}
\usepackage[english]{babel}

\definecolor{sugg}{rgb}{0,0,0}
\newcommand{\sugg}[1]{#1}

\definecolor{light}{rgb}{0,0,0}
\newcommand{\lt}[1]{ #1}

\definecolor{sug}{rgb}{0,0,0}
\newcommand{\sug}[1]{ #1}

\definecolor{quest}{rgb}{1,0,0}

\usepackage{hyperref}

\newcommand{\er}{~\eqref}

\begin{document}

\title{
\sugg{Chiral Relaxation Time at}
the Crossover of Quantum Chromodynamics}

\author{M. Ruggieri}\email{marco.ruggieri@ucas.ac.cn}
\affiliation{College of Physics, University of Chinese Academy of Sciences, 
Yuquanlu 19A, Beijing 100049, China}

\author{G. X. Peng}\email{gxpeng@ucas.ac.cn}
\affiliation{%
College of Physics, University of Chinese Academy of Sciences, 
Yuquanlu 19A, Beijing 100049, China}
\affiliation{Theoretical Physics Center for Science Facilities, Institute of High Energy Physics, Beijing 100049, China}

\author{M. Chernodub}\email{maxim.chernodub@lmpt.univ-tours.fr}
\affiliation{\sugg{CNRS, Laboratoire de Math\'ematiques et Physique Th\'eorique UMR 7350, Universit\'e de Tours, 37200 France}}
\affiliation{\sugg{Soft Matter Physics Laboratory, Far Eastern Federal University, Sukhanova 8, Vladivostok, Russia}}


\begin{abstract}
We study microscopic processes 
responsible
for chirality flips in the thermal bath of Quantum Chromodynamics 
at finite temperature and zero baryon chemical potential.
We focus
on the temperature range where the
crossover from 
chirally broken phase to quark-gluon plasma takes place, namely $T \simeq (150, 200)$ MeV.
The processes we consider are quark-quark scatterings mediated by collective excitations with 
the quantum number of pions and $\sigma$-meson, 
hence we refer to these processes simply as \sugg{to} one-pion (one-$\sigma$) exchange\sugg{s}. 
We use a Nambu-Jona-Lasinio model to compute equilibrium properties
of the thermal bath, as well as the relevant scattering kernel  to be used in the collision integral
to estimate the 
chiral
relaxation time 
$\tau$. We find  
$\tau\simeq 0.1 \div 1$
fm/c around the chiral crossover. 
\end{abstract}

\pacs{12.38.Aw,12.38.Mh}
\keywords{Quark-gluon plasma, Relaxation time, Chiral chemical potential, Nambu-Jona-Lasinio model, 
Chiral phase transition.} 

\maketitle


\section{Introduction}
Interactions of fermions with nontrivial gauge field configurations 
\sugg{carrying} a finite winding number, $Q_W$, 
\sugg{lead to chiral imbalance between the densities of  right-handed, $n_R$, and left-handed, $n_L$, chiral fermions.}
\sugg{The imbalance}
\sugg{-- induced via}
the Adler-Bell-Jackiw (ABJ) anomaly~\cite{Adler:1969gk,Bell:1969ts}
\sugg{ --  is characterized by the finite chiral density, $n_5\equiv n_R-n_L$.}
In 
\sugg{finite-temperature}
Quantum Chromodynamics (QCD)  
\sugg{the topological}
gauge field configurations 
in Minkowski space are
named sphalerons, whose production rate has been estimated to be quite large~\cite{Moore:2000ara,Moore:2010jd}.
The large number of sphaleron transitions at high temperature
suggests the possibility that net chirality might be abundant (locally) in the quark-gluon plasma phase of QCD.
This 
\sugg{observation}
stimulated many studies 
\sugg{of various}
exotic effects, see~\cite{Kharzeev:2007jp,Fukushima:2008xe,Li:2014bha,
Son:2009tf,Banerjee:2008th,Landsteiner:2011cp,Son:2004tq,
Metlitski:2005pr,Kharzeev:2010gd,Chernodub:2015gxa,Chernodub:2015wxa,Chernodub:2013kya,Braguta:2013loa,
Sadofyev:2010pr,Sadofyev:2010is,Khaidukov:2013sja,Kirilin:2013fqa,Avdoshkin:2014gpa} and references therein.

In order to describe 
\sugg{equilibrium}
systems with 
\sugg{a finite chiral density}
$n_5\neq0$ 
it is customary to introduce the chiral chemical potential, $\mu_5$,
conjugated to 
\sugg{the chiral density}
$n_5$~\cite{Ruggieri:2016lrn,Ruggieri:2016cbq,Ruggieri:2016ejz,Frasca:2016rsi,
Gatto:2011wc,Fukushima:2010fe,Chernodub:2011fr,Ruggieri:2011xc,Yu:2015hym,Yu:2014xoa,
Braguta:2015owi,Braguta:2015zta,Braguta:2016aov,Hanada:2011jb,Xu:2015vna,
Wang:2015tia,Ebert:2016ygm,Afonin:2015cla,Andrianov:2012dj,Farias:2016let}. 
Because of 
\sugg{the chiral} 
ABJ anomaly as well as of chirality changing processes
\sugg{the chiral density}
$n_5$ is not a strictly conserved quantity. 
One might, however, assume that 
\sugg{the chiral chemical potential} $\mu_5\neq0$ describes a system in thermodynamical equilibrium with a fixed value of 
\sugg{the chiral charge}
$n_5$ on a time scale which much larger than 
the typical 
\sugg{chiral relaxation}
time scale 
\sugg{$\tau$ which is}
needed for $n_5$ to equilibrate. For example, this equilibration has been recently studied in~\cite{Ruggieri:2016lrn}
where $n_5$ is generated dynamically 
\sugg{via the chiral anomaly activated by}
the simultaneous presence of parallel electric and magnetic fields.

In this article we 
\sugg{compute the chiral relaxation time}, $\tau$,
in a two flavor Nambu-Jona-Lasinio (NJL) model 
\sugg{which is invoked} 
to mimic the QCD thermal bath 
\sugg{in the temperature range $T \simeq (150,200)$ MeV 
where the crossover from color confinement phase to quark-gluon plasma takes place.}
The main processes we consider are quark-quark scatterings mediated by collective excitations
with the quantum numbers of pions, hence we simply call these processes one-pion exchange. 
We also consider, for completeness, scattering mediated by $\sigma$-meson exchange, which however
is less relevant both because of the larger $\sigma$ mass and because of the smaller weight of
the diagrams  with $\sigma-$exchange compared to the ones with a pion exchange.

We use the NJL model to evaluate the chiral condensate at finite temperature, which allows to compute
the constituent quark mass in the thermal bath. Once the quark mass is known, we use the 
well established NJL formalism to compute the scattering kernel of the microscopic processes we consider,
and the collision integral (which represents the main numerical computation of this work) that allows to
estimate the relaxation time of chirality.
\sugg{The main result of our article is that}
we find 
\sugg{$\tau \simeq 0.1 \div 1$} fm in the aforementioned crossover temperature range.
We also find that the relaxation time decreases with temperature, regardless of the fact that
chiral symmetry gets partially restored at the crossover. This is explained taking into account 
that although the scattering kernel of chirality flipping processes becomes smaller with increasing temperature,
the portion of phase space for the scattering increases with temperature, eventually leading to
an increase of the scattering rate and a lowering of the relaxation time.
 The behavior of $\tau$ versus temperature computed here is in 
\sugg{some disagreement} with the ansatz
$\tau \propto 1/M_q$ used in Ref.~\cite{Ruggieri:2016lrn}, where $M_q$ corresponds to the
constituent quark mass, which 
\sugg{was} admittedly too simple as it did not take into 
account properly the phase space opening at finite temperature.

We also consider a computation in which we merge by hand 
the NJL model below the critical temperature
with a quasiparticle model above the critical temperature:
the latter differs from NJL because the quasiparticle thermal mass,
that is obtained by a numerical fit of Lattice data about pressure, entropy and energy density, 
is assumed to arise as a pole in the propagator dressed with 
a chiral invariant self-energy rather than from a 
term $\propto\bar\psi\psi$ in the lagrangian. This thermal mass is generated by 
many body effects that are not present in a mean field NJL: in fact the latter decribes
only the constituent mass arising from both spontaneous and explicit chiral symmetry breaking. 
In quasiparticle models the thermal mass is generally
found to be large around the chiral crossover and increasing with temperature
\cite{Gorenstein:1995vm,Plumari:2011mk,Levai:1997yx,Bluhm:2004xn,Bluhm:2007nu,Castorina:2011ja,Alba:2014lda,Ruggieri:2012ny,Oliva:2013zda,Begun:2016lgx};
this large mass suppresses thermal quark excitations, therefore we expect that  
interpolating between NJL and quasiparticles around the chiral crossover
will lower the collision rate, hence increase the relaxation time.
This rough idea is in agreement with our results.
We however do not push very much the results of this calculation since it is far from being rigorous:
our only purpose is to illustrate how the collision rate would be affected
if beside the NJL constituent quark mass one introduces in an effective way 
many body effects encoded in a large thermal mass above $T_c$.

The plan of the article is as follows. In Section II we state the problem and set up the main equations 
needed to compute the relaxation time. 
In Section III we discuss the one-meson exchange within the NJL model. 
In Section IV we summarize our main findings, in particular the relaxation time shown in
Figg.~4 and~8.. 
Finally in Section V we draw our Conclusions.

\section{Relaxation time of chiral density}
\sugg{The main aim of this article is to compute
the relaxation time $\tau$ for chirality flips $R \leftrightarrow L$ in a thermal bath at given temperature temperature $T$.}
\sugg{In our effective-model approach the specific microscopic process responsible for the the chirality flips is to pion-exchange between left- and right-handed quarks. We will discuss this process in details below.}
In this Section we begin with a brief statement of the problem
of chiral density relaxation, then we define the microscopic process 
\sugg{that change}
chirality in the thermal bath and 
\sugg{then we}
compute the relevant scattering matrix.   

In order to state the problem we consider quark matter
in the background of parallel electric and magnetic fields~\cite{Ruggieri:2016lrn,
Warringa:2012bq,Babansky:1997zh,Cao:2015cka}. In this case
the evolution of the chiral density 
\sugg{$n_5$ with time} is given by
\begin{equation}
\frac{dn_5}{dt} =- n_5 \Gamma + N_c\frac{ (eE)(eB)}{2\pi^2}\sum_f q_f^2
e^{-\frac{\pi M^2}{|q_f eE|}},
\label{eq:gamma}
\end{equation}
where the first term on the right hand side 
\sugg{describes a relaxation of the chiral density due to}
chirality-changing processes in the thermal medium, while the second term comes from the ABJ 
\sugg{chiral}
anomaly 
\sugg{supplemented with the exponential prefactor which takes into account the finiteness of the quark mass $M$}.
\sugg{The quantity}
$\Gamma$ in Eq.\er{eq:gamma}
corresponds to the rate of the chirality 
\sugg{flips}
while its inverse defines the 
\sugg{chiral relaxation time},
\begin{equation}
\tau = 1/\Gamma\,.
\label{eq:tau}
\end{equation}

\sugg{Physically,
Eq.~\er{eq:gamma} indicates that in parallel electric and magnetic external fields the ABJ anomaly creates a chiral imbalance $n_5$.  According to the second term of Eq.~\er{eq:gamma}, the chiral density should start grow linearly with time if even the chiral imbalance was initially absent in the system, $n_5 = 0$. This process would continue forever -- as long as the external fields  are not screened by the media -- if there were no other processes in the system. However, in the thermal bath certain microscopic processes may flip the chirality of quarks and the significance of the chirality-flipping process increases with the increase of the chiral density $n_5$. These processes are encoded in the first term in Eq.~\er{eq:gamma} where $\Gamma$ is the chirality-changing rate which defines the characteristic chiral relaxation time $\tau$, Eq.~\er{eq:tau}. If one waits long enough, $t\gg\tau$, the value of chiral density $n_5$ exponentially equilibrates at the following value:}
\begin{equation}
n_5^{\mathrm{eq}} =N_c \frac{(eE)(eB)}{2\pi^2}\tau\sum_f   q_f^2 
e^{-\frac{\pi M^2}{|q_f eE|}}.
\end{equation}
The knowledge of the relaxation time $\tau$ is therefore crucial since it allows to determine the equilibrium value of chiral density, $n_{5}^{\mathrm{eq}}$, and to compute the 
\sugg{thermodynamically} 
conjugated chiral chemical potential $\mu_5$.

As we describe in more details below, the microscopic processes we are interested in are 
quark-quark scattering mediated by collective modes with the quantum numbers of pions,
hence we refer to these processes as one-pion exchange for simplicity. 
The computation of $\tau$ for the physical 
setup described above -- i.e. for the system of quarks in external electromagnetic fields --
is too complicated because the external fields 
\sugg{create a finite chiral density $n_5$ which is associated with nonzero chiral chemical potential $\mu_5$}; 
\sugg{the nonzero $\mu_5$ and the fields affect the quark propagators,}
indirectly changing meson properties and scattering amplitudes,
making the consistent calculation very tough.
\sugg{Therefore,} 
here we limit ourselves to a much simpler problem, namely to the computation of 
\sugg{relaxation time}
$\tau$ for a system without external fields 
\sugg{at negligibly small value of the chiral chemical potential $\mu_5 \ll T$.}
\sug{In this paper we are not interested in concrete physical mechanism which creates the chiral imbalance.}
Similar approach has already been used in~\cite{Manuel:2015zpa} to estimate $\tau$ in quark-gluon plasma, 
where 
\sugg{gluon- and photon-mediated}
Compton scattering processes have been taken as the microscopic mechanisms for chirality flips in the thermal bath. 
\sug{Here we differ from Ref.~\cite{Manuel:2015zpa} because in the region of the chiral crossover the pion-exchange processes are much more effective compared to the Compton scattering.\footnote{\sug{Using explicit expressions of Ref.~\cite{Manuel:2015zpa} we estimated that the pion-mediated processes -- with the relaxation time given in our Fig.~\ref{Fig:4hjklm} below -- are about one- or even two-order to magnitude faster compared to the Compton scattering.}}}

By the very definition of the chiral density,
\begin{equation}
n_5 = N_c N_f\int\frac{d^3 p}{(2\pi)^3}\left( f_R  -  f_L \right),
\label{eq:PS1a}
\end{equation}
where $f_{R,L}$ denote distribution functions for the 
\sugg{right-handed and left-handed}
quarks respectively, we get
\begin{equation}
\frac{d n_5}{d t} = N_c N_f\int\frac{d^3 p}{(2\pi)^3}\left(\frac{d f_R}{dt} - \frac{d f_L}{dt} \right).
\label{eq:PS1}
\end{equation}
 The overall $N_c N_f$
takes into account that $n_5$ is defined as a sum over color and flavor; keeping this in mind, $f_{R,L}$ denote
distribution functions for a quark with color and flavor fixed.
Time evolution of $f_{R,L}$ is given by the Boltzmann collision integral,
\begin{eqnarray}
\frac{d f_R(p)}{dt} &=& \int d\Pi\frac{(2\pi)^4\delta^4(p+k-p^\prime-k^\prime)}
{2E_p}|{\cal M}|^2 F, 
\label{eq:PS2}
\end{eqnarray}
where $d\Pi$ corresponds to 
\sug{the standard}
momentum space measure,
\begin{equation}
d\Pi = \frac{d^3k}{(2\pi)^3 2E_k}
 \frac{d^3k^\prime}{(2\pi)^3 2E_k^\prime}
  \frac{d^3p^\prime}{(2\pi)^3 2E_p^\prime},
\end{equation}
and \sug{the kernel $F$ takes into account the population of the incoming and outgoing particles in the process.}
In Eq.\er{eq:PS2} the squared transition amplitude $|{\cal M}|^2$ is the main ingredient in the collision 
integral and it can be computed once a microscopic process has been chosen.

The microscopic processes on which we focus in this article are 
\sug{the transitions} $q_R q_R \rightarrow q_L q_L$ and vice-versa.
The change of chiral density produced by these processes would cancel if in the thermal bath \sug{is chirally balanced,} $\mu_5 = 0$.
However the presence \sug{of the chiral imbalance} $\mu_5\neq 0$ implies a different population of $R$- and
$L$-handed quarks, which in turn results in a finite rate for the chiral density equilibration. 
Given this microscopic process it is possible to specify the \sug{kernel} $F(p,k,p^\prime,k^\prime)$ in Eq.\er{eq:PS2}.
In this study we consider the simple case of the classical Boltzmann kernel, namely
\begin{eqnarray}
F(p,k,p^\prime,k^\prime)&=&f_L(p^\prime)f_L(k^\prime)-f_R(p)f_R(k)
 \label{eq:op1}
\end{eqnarray}
for the scattering of two incoming $R$ quarks giving two outgoing $L$ quarks.
\sug{The Boltzmann} distribution functions are defined as
\begin{equation}
f_{R/L}(p)=e^{-\beta\omega_\pm},
\end{equation}
\sug{where the dispersion relation is}
\begin{equation}
\omega_s = \sqrt{(p+s\mu_5)^2 + M_q^2},~~~~~s=\pm 1.
\label{eq:omS}
\end{equation} 
We also consider the Fermi-Dirac kernel,
\begin{eqnarray}
F(p,k,p^\prime,k^\prime)&=&f_L(p^\prime)f_L(k^\prime)[1-f_R(p)][1-f_R(k)]
-f_R(p)f_R(k)[1-f_L(p^\prime)][1-f_L(k^\prime)],
\label{eq:op2}
\end{eqnarray}
\sug{where the distribution functions} 
\begin{equation}
f_{R/L}(p)= \frac{1}{1 + e^{\beta\omega_{\pm}}},
\end{equation}
\sug{correctly take into account the Pauli blocking} due to the fermionic nature of quarks.

For simplicity we limit ourselves to consider the lowest order in $\mu_5/T$ in the collision integral\sug{\er{eq:PS2}}. 
When we combine $d f_R/dt$ and $d f_L/dt$ in Eq.\er{eq:PS1} we take into account that 
\begin{equation}
\frac{d f_L}{dt} = \frac{d f_R}{dt}(\mu_5\rightarrow -\mu_5),
\label{eq:ex}
\end{equation}
so only the odd part in $\mu_5$ contributes to $dn_5/dt$. 
It is easy to verify that both the Dirac delta argument and the four energies in the denominator
are even functions of $\mu_5$, thus it is enough to consider these at $\mu_5=0$ and expand
$F(p,k,p^\prime,k^\prime)$ in Eq.\er{eq:op1} up to the first order in $\mu_5/T$. Writing
\begin{equation}
\frac{d f_R}{dt} = {\cal A}(p) + \mu_5 {\cal B}(p) + O(\mu_5^2)\label{eq:defin}
\end{equation}
and taking into account Eq.\er{eq:ex} we have
\begin{equation}
\frac{dn_5}{dt} = 2N_c N_f \mu_5 \int\frac{d^3 p}{(2\pi)^3}{\cal B}(p),
\label{eq:aplp}
\end{equation}
where we have taken into account the color-flavor degeneracy; 
the squared matrix element to use in Eq.\er{eq:aplp} is given by 
Eq.\er{eq:msqEEE}.
The collision rate for chirality change, $\Gamma$, is obtained from Eq.\er{eq:gamma}, namely
\begin{equation}
\Gamma=-\frac{1}{n_5}\frac{dn_5}{dt},\label{eq:GAMMAghj}
\end{equation}
and the relaxation time is then computed by Eq.\er{eq:tau}.

In order to relate the chiral density to the chiral chemical potential we use the NJL model at finite 
$\mu_5$~ \cite{Fukushima:2010fe,Gatto:2011wc},
limiting ourselves to the leading order in $\mu_5/T$. \sug{The chiral density} $n_5$ can be computed
as $n_5 = -\partial\Omega/\partial \mu_5$ where $\Omega$ is the thermodynamic
potential, 
\begin{equation}
\Omega =\Omega_{MF} + \Omega_v + \Omega_T,
\label{eq:apo1}
\end{equation}
with
\begin{eqnarray}
\Omega_{MF} &=& \frac{(M_q - m_0)^2}{4G},\\
\Omega_v &=&  -N_c N_f\sum_{s=\pm 1} \int\frac{d^3 p}{(2\pi)^3}\omega_s,\\
\Omega_T &=& -2N_c N_f T \sum_{s=\pm 1} \int\frac{d^3 p}{(2\pi)^3}
\log\left(1+e^{-\beta\omega_s}\right).
\end{eqnarray}
In the above equations $M_q = m_0 -2G\langle\bar{q}q\rangle$
with $\langle\bar{q}q\rangle = \langle\bar{u}u\rangle + \langle\bar{d}d\rangle$
corresponding to the chiral condensate.
To obtain the chiral density as a function of $\mu_5$ we expand Eq.\er{eq:apo1} up to $O(\mu_5^2 T^2)$:
\begin{equation}
\Omega = \Omega_0 + \mu_5^2 \left(L_0+L_T\right),
\label{eq:Omega:2}
\end{equation}
where $\Omega_0$ corresponds to the thermodynamic potential for $\mu_5=0$, namely
\begin{eqnarray}
\Omega_0 &=& \frac{(M_q - m_0)^2}{4G} - 2N_c N_f\int\frac{d^3p}{(2\pi)^3}\omega
-4N_c N_f T \int\frac{d^3p}{(2\pi)^3} \log\left(1+e^{-\beta \omega}\right),
\label{eq:NJL3}
\end{eqnarray}
with $\omega = \sqrt{p^2 + M_q^2}$. 

\sug{The term quadratic in $\mu_5$ in Eq.\er{eq:Omega:2} comes with the prefactors}:
\begin{eqnarray}
L_0 &=&-\frac{N_c N_f}{2\pi^2} M_q^2 \int_0^\Lambda dp 
\frac{p^2}{(p^2 + M_q^2)^{3/2}},\\
L_T &=& -\frac{N_c N_f}{\pi^2}
\int_0^\infty dp \frac{p^2}{(p^2 + M_q^2)^{3/2}} 
\frac{ \left(-M_q^2
   e^{\beta  \omega}+\beta 
   p^2 \omega e^{\beta 
   \omega}-M_q^2\right)}{ 
    \left(e^{\beta 
   \omega}+1\right)^2}.
\end{eqnarray}
By virtue of the equations above we can write 
\begin{equation}
n_5 = -2\mu_5(L_0 + L_T).\label{eq:n5tyu}
\end{equation}
It can be easily verified that $L_0$ vanishes for $M_q=0$. On the other hand in the limit of vanishing quark mass 
the above equation leads to $n_5 = N_c N_f \mu_5 T^2/3$ in agreement with Ref.~\cite{Fukushima:2008xe}.  

Divergent ultraviolet integrals in the above equations are regulated
by a hard cutoff $\Lambda$, where $\Lambda$ is considered as one
of the parameters of the model and its value is fixed by phenomenological requirements together with
the NJL coupling, $G$, and the bare quark mass. The parameter set we use is
$\Lambda = 653$ MeV, $m_0 = 5.39$ MeV and $G=2.14/\Lambda^2$.
Taking into account Eqs.\er{eq:aplp},\er{eq:GAMMAghj} and\er{eq:n5tyu} we can write the rate for the chirality change as
\begin{equation}
\Gamma=\frac{N_c N_f}{L_0+L_T}\int\frac{d^3 p}{(2\pi)^3}{\cal B}(p),
\label{eq:GAMMAequ}
\end{equation}
with ${\cal B}(p)$ defined in Eq.\er{eq:defin}. The relaxation time for chirality is then given by Eq.\er{eq:tau}.

\section{One-meson exchange within the NJL model}
We are interested to interaction channels which lead to a change of chiral density in the thermal bath.
Here we focus on one pion exchange, which should be the dominant process around the chiral crossover. 
We also consider scattering mediated by $\sigma$ meson but its contribution to the collision rate is found
to be smaller than the one obtained by one-pion exchange.
We assume 
\sug{that}
quarks are in equilibrium in a thermal bath with temperature $T$ and chiral chemical potential 
$\mu_5\neq0$ (a vanishing $\mu_5$ would lead to a zero net chiral density change
by this process) and we focus on transitions $q_R q_R\rightarrow q_L q_L$ and 
$q_L q_L\rightarrow q_R q_R$. \sug{Strictly speaking, a system with the chiral imbalance, $\mu_5 \neq 0$, cannot be in thermal equilibrium due to chirality-changing processes which are the subject of this article. However, we assume that there is a process that pumps the chiral charge into the system so that the mean chiral density and, consequently, the chiral chemical potential, are both nonzero. The chiral charge may be pumped into the system by the chiral anomaly in the background of parallel electric and magnetic fields 
(see, e.g., Ref.~\cite{Ruggieri:2016lrn} for the relevant discussion in the context of the NJL model).}

\subsection{Quark-pion scattering kernel}
In order to compute the rate for chirality changing processes\er{eq:tau} in the medium close to the
chiral phase transition we use a two flavor Nambu-Jona-Lasinio (NJL) model~\cite{Nambu:1961tp,Nambu:1961fr}
(see~\cite{Klevansky:1992qe,Hatsuda:1994pi} for reviews) with lagrangian density given by
\begin{equation}
{\cal L} =\bar q \left(i\gamma_\mu \partial^\mu - m_0\right) q + {\cal L}_{4},
\label{eq:NJL1}
\end{equation}
where $q$ denotes a quark field with Dirac, color and flavor indices and $m_0$ is the current
quark mass. In the above equation the interaction lagrangian, ${\cal L}_{4}$, is given by
\begin{equation}
{\cal L}_{4} = G\left[(\bar q q)^2 + (\bar q i \gamma_5 \bm\tau q)^2\right],
\end{equation}
which is invariant under $SU(2)_V \otimes SU(2)_A \otimes U(1)_V$ \sug{group}, and $G$ is a coupling
constant with mass dimension $d=-2$. Introducing the collective fields
$\sigma=G\bar q q$, $\bm\pi = G\bar q i\gamma_5\bm\tau q$, after a 
Hubbard-Stratonovich transformation
\sug{the interaction term}
${\cal L}_4$ can be written as 
\begin{equation}
{\cal L}_{4} = 2G\langle\bar{q}q\rangle \bar{q}q + \bar q \left[
g_{\sigma qq}^0 \sigma + g_{\pi qq}^0 i \gamma_5 \bm\tau\cdot\bm \pi 
\right] q  - \frac{(G\langle\bar{q}q\rangle+\sigma)^2 + \bm\pi^2}{G},
\label{eq:L4}
\end{equation}
where we have introduced the bare quark-meson couplings $g_{\sigma qq}^0 = g_{\pi qq}^0 = 2$. The bare couplings get renormalized by quark interactions in the medium 
\sug{and they} 
give effective quark-meson couplings. From now on we denote by $\sigma$ the quantum fluctuation
of the collective field $G\bar q q$ on the top of its expectation value $G\langle\bar q q\rangle$.
In the partition function of the model specified by Lagrangian\er{eq:L4} an integration over
quark and meson fields is understood; the Lagrangian is quadratic in quark fields so the functional
integral can be done exactly and one is left with an effective Lagrangian for meson fields,
whose in-medium propagators can be computed easily by the random phase approximation;
it is then possible to write an effective quark-quark interaction due to one meson exchange: 
\sug{since this topic is well established in the literature, below we quote, without derivations, the basic equations relevant for the present study and refer the interested reader to the review~\cite{Klevansky:1992qe} for further details.}
We firstly focus on the one-pion exchange as it will be the dominant process in the temperature range
of our interest; the description of $\sigma-$meson exchange can be obtained easily once the formalism
for the pion exchange is established.

From Eq.\er{eq:L4} we can extract the quark-pion interaction at the tree level,
\begin{eqnarray}
{\cal L}_{\pi qq} &=& i  g_{\pi qq}^0 \bar q   \gamma_5 \bm\tau\cdot\bm \pi q 
= i  g_{\pi qq}^0 \left(
\bar q_L \bm\tau\cdot\bm \pi q_R  -   \bar q_R \bm\tau\cdot\bm \pi q_L 
\right), 
\label{eq:pqqTL}
\end{eqnarray}
where we have made explicit the change of chirality of quarks due to the interaction
with a pion-like collective excitation. Therefore it is possible to change the net chirality of the system
by virtue of processes $q_R q_R \rightarrow q_L q_L$ and viceversa.
The change of chiral density produced by these processes would cancel if in the thermal bath $\mu_5 = 0$;
however assuming a $\mu_5\neq 0$ implies a different population of $R$- and
$L$-handed quarks, which in turn results in a finite rate for chiral density equilibration
\sug{given in \er{eq:GAMMAequ} and\er{eq:tau}}.

The amplitude for the scattering process $qq\rightarrow qq$ due to one pion exchange can be written as
\begin{equation}
i{\cal M} = i\bar q_{ai} q_{bj} \bar q_{ck} q_{d\ell} (U_{\alpha\beta})_{ijk\ell}^{abcd},
\label{eq:ghjk}
\end{equation}
where $a,\dots,d$ denote Dirac indices and $i,\dots,\ell$ correspond to flavor indices
(one meson exchange is
blind to color hence there is no need to introduce a color index in the above equation). The scattering kernel,
$U$, is given in the random phase approximation by
\begin{equation}
i (U_{\alpha\beta})_{ijk\ell}^{abcd} =i ({\cal T}_\alpha)_{ij}^{ab}
\frac{2G}{1 - 2 G \Pi}
({\cal T}_\beta)_{k\ell}^{cd},\label{eq:rpa1}
\end{equation}
where interaction vertex ${\cal T}$ carries Dirac and flavor structure and depends on the particular
interaction channel:
\begin{eqnarray}
{\cal T}_\alpha &=& i \gamma_5\otimes T_{\alpha}.
\end{eqnarray}
For $\pi_0$ exchange ${\cal T}_\alpha = {\cal T}_\beta = \sigma_3$ with $\sigma_3$ being the
third Pauli matrix in flavor space; for $\pi^{\pm}$ exchange one has to use the combinations
\begin{equation}
\tau^\pm = \frac{1}{\sqrt{2}}\left(\sigma_1 \pm \sigma_2\right).
\end{equation}
However the scattering amplitude does not depend on the particular channel chosen among the neutral
and charged pion exchanges (neglecting the small mass difference between $\pi^\pm$ and $\pi_0$), 
therefore from now on we suppress the greek indices and we focus on $\pi_0$ exchange. 
From Eq.\er{eq:rpa1} it is possible to read the in-medium meson propagator
in momentum space,
\begin{equation}
D(k^2) = \frac{2 G}{1 - 2 G \Pi(k^2)},
\label{eq:mpp}
\end{equation}
where the pion self-energy is given by
\begin{equation}
\Pi(k^2) = -i\text{Tr}\int\frac{d^4p}{(2\pi)^4}
\gamma_5 \sigma_3 S(p) \gamma_5 \sigma_3 S(p-k).
\end{equation}
A standard algebraic manipulation leads to~\cite{Klevansky:1992qe}
\begin{equation}
\Pi(k_0,\bm k) = \frac{1}{2G}\left(1-\frac{m_0}{M_q}\right)
+ 2 N_c N_f k^2 I(k^2),\label{eq:pi1}
\end{equation} 
where\footnote{Our definition of $I$ differs from the oen of~\cite{Klevansky:1992qe} for an overall $-i$.}
\begin{equation}
I(k_0, \bm k) = -i\int\frac{d^4p}{(2\pi)^4}\frac{1}{(p^2 - M_q^2)[(p-k)^2 - M_q^2]}
\label{eq:IIIp}
\end{equation}
and $k^\mu=(k_0,\bm k)$ on the right hand side of the above equation.

\subsection{$\sigma$-quark scattering kernel}
The formalism set up in the previous section for the quark-pion scattering
can be adapted easily to the description of the scattering kernel of quarks with $\sigma$-meson.
In particular, the amplitude for the scattering process $qq\rightarrow qq$ due to one $\sigma$ exchange can be written 
analogously to Eq.\er{eq:ghjk}, namely
\begin{equation}
i{\cal M} = i\bar q_{ai} q_{bj} \bar q_{ck} q_{d\ell} (U)_{ijk\ell}^{abcd},
\label{eq:ghjk2}
\end{equation}
with the scattering kernel within the random phase approximation given by
\begin{equation}
i (U)_{ijk\ell}^{abcd} =i ({\cal T})_{ij}^{ab}
\frac{2G}{1 - 2 G \Pi_\sigma}
({\cal T})_{k\ell}^{cd},\label{eq:rpa456}
\end{equation}
and
\begin{eqnarray}
{\cal T} &=& \bm 1_D\otimes \bm 1_F,
\end{eqnarray}
where $1_D$ {\sug{and $1_F$}} denote the identity in Dirac and flavor spaces, respectively.
From Eq.\er{eq:rpa456} it is possible to read the in-medium $\sigma$ propagator
in momentum space,
\begin{equation}
D_\sigma(k^2) = \frac{2 G}{1 - 2 G \Pi_\sigma(k^2)},
\label{eq:mppSS}
\end{equation}
with self-energy given by
\begin{equation}
\Pi_\sigma (k^2) = -i\text{Tr}\int\frac{d^4p}{(2\pi)^4}
\gamma_5 \sigma_3 S(p) \gamma_5 \sigma_3 S(p-k).
\end{equation}
A standard algebraic manipulation leads to~\cite{Klevansky:1992qe}
\begin{equation}
\Pi_\sigma(k_0,\bm k) = \frac{1}{2G}\left(1-\frac{m_0}{M_q}\right)
+ 2 N_c N_f (k^2 - 4 M_q^2) I(k^2),\label{eq:pi1SS}
\end{equation} 
with $I$ defined in Eq.\er{eq:IIIp}.

\subsection{Scattering amplitude: $\pi$-exchange}

\begin{figure*}[t!]
\begin{center}
\includegraphics[width=16cm]{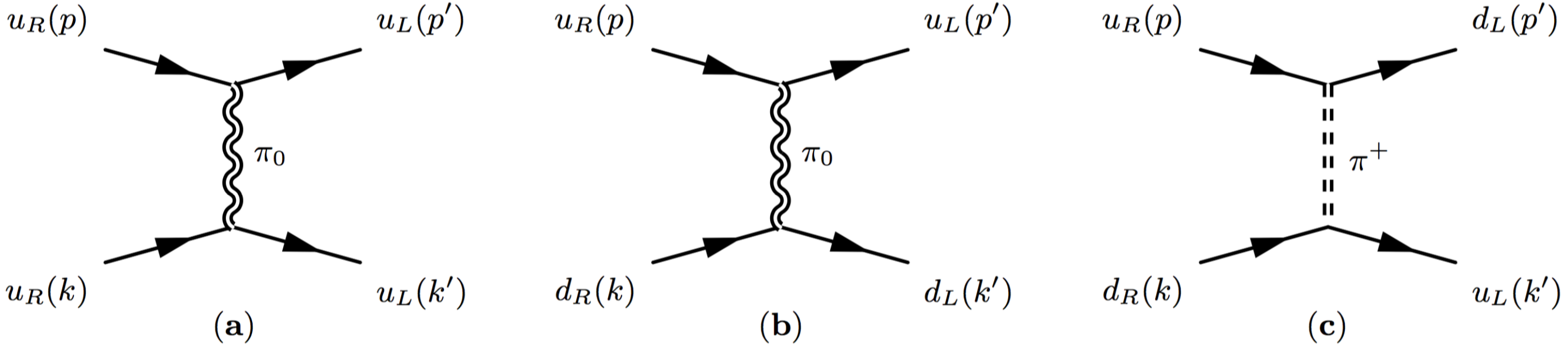}
\end{center}
\caption{\label{Fig:diagra}Tree level diagrams for the chirality flips of $u$ quarks due to one-pion exchange.}
\end{figure*}

By means of the one pion exchange it is possible to write several diagrams 
giving contribution to the scattering amplitude, represented in Fig.~\ref{Fig:diagra}
for the case of an incoming $u_R$ quark.
We assume equal mass for charged and neutral pions, and $\mu_{5u} = \mu_{5d}$:
in this way the one pion exchange is blind to quark flavor and the scattering amplitude is independent
on the particular current chosen. 
Given a quark with color and flavor fixed, for the $\pi_0$ exchange there are
two possible processes, namely $\mathbf{(a)}$ and $\mathbf{(b)}$ in Fig.~\ref{Fig:diagra},
and the charged pion exchange adds one further process denoted by $\mathbf{(c)}$ in Fig.~\ref{Fig:diagra}.
In this isospin symmetric limit the three diagrams in Fig.~\ref{Fig:diagra} give the same result.
We denote by ${\cal M}_i$ the antisimmetrized amplitude corresponding to diagram $\mathbf{(i)}$
with $i=a,b,c$.
For each of the diagrams with incoming $u_R$ quark in Fig.~\ref{Fig:diagra} 
we sum incoherently on the color of the second incoming quark (namely, cross sections are summed
rather than amplitudes): this brings an overall $N_c$ to the total cross section. 
In addition we sum incoherently over flavors, considering however that
diagrams $\mathbf{(b)}$ and $\mathbf{(c)}$ correspond to the same initial and final states so the corresponding
amplitudes should add coherently. Therefore we can write the squared amplitude as
\begin{equation}
|{\cal M}|^2 = N_c |{\cal M}_a|^2 + N_c   |{\cal M}_b + {\cal M}_c|^2.
\end{equation}
\sug{Since} in the isospin symmetric limit \sug{the amplitude} ${\cal M}_i$ does not depend on the index $i$ we can write
\begin{equation}
|{\cal M}|^2 = N_c  (1+4) |{\cal M}_a|^2.
\label{eq:mtotAAA} 
\end{equation}

The calculation of the transition amplitude is quite standard: the only detail to take into account is the
projection of initial and final states onto chirality eigenstates. This is achieved noticing that
the current can be written as
$\bar{q}_L\gamma_5 q_R = \bar{q}\gamma_5 P_R q$,
where $P_R=(1+\gamma_5)/2$ and a similar relation holds for the current changing an incoming left to an outgoing right.
Therefore we can use all the standard technology for tree level calculations of transition amplitudes
forgetting the selection of chirality in the spinors, because it  is automatically implemented
thanks to the chirality projector.
In the isospin symmetric limit we {\sug{find that}} the diagrams in Fig.~\ref{Fig:diagra} give the same contribution.
We obtain:
\begin{eqnarray}
{\cal M}_{a} = D(t)
\left(
\bar u_{p^\prime}^\prime\gamma_5 P_R u_p
\right)
\left(
\bar u_{k^\prime}^\prime\gamma_5 P_R u_k
\right) 
-D(u)
\left(
\bar u_{k^\prime}^\prime\gamma_5 P_R u_1
\right)
\left(
\bar u_{p^\prime}^\prime\gamma_5 P_R u_2
\right);\label{eq:1}
\end{eqnarray}
here the prime denotes outgoing particles, and labels $1$, $2$ label the particle, 
$u$ and $t$ denote standard Mandelstam variables.
In writing Eq.\er{eq:1} we have ignored all the
overall $i$ since they do not affect the squared matrix element. 
Taking into account that for each color of the incoming quark $p$ there are
$N_c$ possible colors of the incoming $k$ for the scattering, and that the scattering
involving different colors sum up incoherently, we can write the squared matrix element as
\begin{eqnarray}
\left|{\cal M}\right|_{\pi qq}^2 = \frac{1}{4}\left[
4N_c(1+4)D(t)D^\dagger(t) a_1
+4N_c(1+4)D(u)D^\dagger(u)a_2 -2 N_c (1+4)D_{tu}
\left(a_1 + a_2- a_3\right)
\right],
\label{eq:msqEEE}
\end{eqnarray}
with $D_{tu} =D(t)D^\dagger(u) + D(u)D^\dagger(t)$ and $a_i$ defined as
\begin{eqnarray}
a_1 &=& \frac{(t-2M_q)^2}{4},\\
a_2 &=&\frac{(u-2M_q)^2}{4},\\
a_3 &=&\frac{(s-2M_q)^2}{4},
\end{eqnarray}
with $t$, $u$ and $s$ denoting the Mandelstam variables. The overall \sug{factor} $1/4$ in Eq.\er{eq:msqEEE}
takes into account the average over initial spins and sum over final spins.

From Eq.\er{eq:msqEEE} we notice that
the scattering takes place in the $t$ and $u$ channels: because $t\leq 0$ and $u\leq 0$
the in-medium pion propagator is probed by spacelike virtual momenta and the pion self-energy
$\Pi(k_0,\bm k)$, which is the main ingredient in the scattering kernel
in Eq.\er{eq:rpa1}, has to be computed for $k^2 \leq 0$. 
After analytic continuation to imaginary time and using the Matsubara formalism to deal
with loop integrals at finite temperature we find 
\begin{equation}
I(k_0,\bm k) = -\int\frac{d^3p}{(2\pi)^3}
\left[\frac{A}{k_0 + E_p + E_{pk}} + \frac{B}{k_0 - E_p - E_{pk}}\right],
\label{eq:A1}
\end{equation}
where we have defined 
\begin{eqnarray}
E_p = \sqrt{p^2 + M_q^2}\,, \qquad
E_{pk} = \sqrt{(\bm p - \bm k)^2 + M_q^2}
\end{eqnarray}
and
\begin{eqnarray}
A &=& \frac{1}{4E_p}\frac{\tanh(E_p/(2\beta))}{k_0 + E_p - E_{pk}} + 
\frac{1}{4E_{pk}}\frac{\tanh(E_{pk}/(2\beta))}{k_0 - E_p + E_{pk}} ,
\label{eq:A2}\\
B &=& \frac{1}{4E_p}\frac{\tanh(E_p/(2\beta))}{k_0 - E_p + E_{pk}} + 
\frac{1}{4E_{pk}}\frac{\tanh(E_{pk}/(2\beta))}{k_0 + E_p - E_{pk}}.
\label{eq:A3}
\end{eqnarray}
In Eq.\er{eq:A1} we have made explicit the poles at $k_0 = \pm (E_p + E_{pk})$ which, once treated by the
$i\varepsilon$ prescription to build a Feynman propagator, are responsible for the
pion instability towards creation of quark-antiquark pairs; 
in a similar way in Eqs.\er{eq:A2} and\er{eq:A2}
we have split the contributions in terms of functions that become singular when $k_0 = \pm (E_p - E_{kp})$
that still give an imaginary part and are related to pion emission and absorption processes by quarks.

In the processes of interest in the present article only the latter poles are relevant: as a matter of fact
being $k^2 \leq 0$ implies $k_0^2 \leq \bm k^2$: it is easy to realize
that this condition forces $E_p + E_{pk} > |k_0|$ for any value of $k_0$, hence removing the singularities 
$\propto [k_0 \pm(E_p + E_{pk})]^{-1}$ from the $\bm p$ space. Physically this means that 
the only contribution of the imaginary part of the scattering kernel is related
to emission and absorption processes.
Taking into account only the emission-absorption poles the real and imaginary parts of $I$ in Eq.\er{eq:A1} 
can be easily obtained: treating the poles by the standard $i\varepsilon$ prescription to build a Feynman propagator 
and using of the Sokhotski-Plemelj theorem, 
\begin{equation}
\frac{1}{x-x_0\pm i\varepsilon} = \mp i \pi \delta(x-x_0) + \mathrm{PV}\frac{1}{x-x_0},
\end{equation}
where PV corresponds to the principal value, we find
\begin{eqnarray}
\Im I(k_0,\bm k) &=& \mathrm{sign}(k_0)\pi \int\frac{d^3p}{(2\pi)^3}
\left[
\frac{1}{4E_p}\frac{\tanh(E_p/(2\beta))}{k_0 + E_p + E_{pk}} + 
\frac{1}{4E_{pk}}\frac{\tanh(E_{pk}/(2\beta))}{k_0 - E_p - E_{pk}}
\right]\delta(k_0 + E_p - E_{pk}) \nonumber\\
&&+\mathrm{sign}(k_0)\pi \int\frac{d^3p}{(2\pi)^3}
\left[
\frac{1}{4E_p}\frac{\tanh(E_p/(2\beta))}{k_0 - E_p - E_{pk}} + 
\frac{1}{4E_{pk}}\frac{\tanh(E_{pk}/(2\beta))}{k_0 + E_p + E_{pk}}
\right]\delta(k_0 - E_p + E_{pk}).\label{eq:lop1}
\end{eqnarray}

We can resolve easily the two $\delta$ functions in Eq.\er{eq:lop1}
\begin{eqnarray}
\Im I(k_0,\bm k) &=& \mathrm{sign}(k_0)\pi \int\frac{d^3p}{(2\pi)^3}
\frac{1}{|g^\prime(P_\pm)|}
\left[\frac{\tanh(E_p/(2\beta))}{4E_p}
 +\frac{\tanh(E_{pk}/(2\beta))}{4E_{pk}} \right]
\nonumber\\
&&\times\left(
\frac{1}{k_0 - E_p - E_{pk}} + \frac{1}{k_0 + E_p + E_{pk}}
\right)\left(\delta(p_x - P_+) + \delta(p_x  - P_-)\right),
\end{eqnarray}
where $P_{\pm}$ are the two solutions of the equation
\begin{equation}
g(p_x) \equiv k_0 + E_p - E_{pk}=0,\label{eq:PMM}
\end{equation} 
satisfying $P_- = - P_+$, and $g^\prime$ denotes the derivative of $g$ with respect to $p_x$, 
whose absolute value can be easily proved to be independent on the sign of $p_x$. 
The real part of $I$ is then obtained by taking the principal value of Eq.\er{eq:A1}.

\subsection{Scattering amplitude: $\sigma$-exchange}

\begin{figure*}[t!]
\begin{center}
\includegraphics[width=12cm]{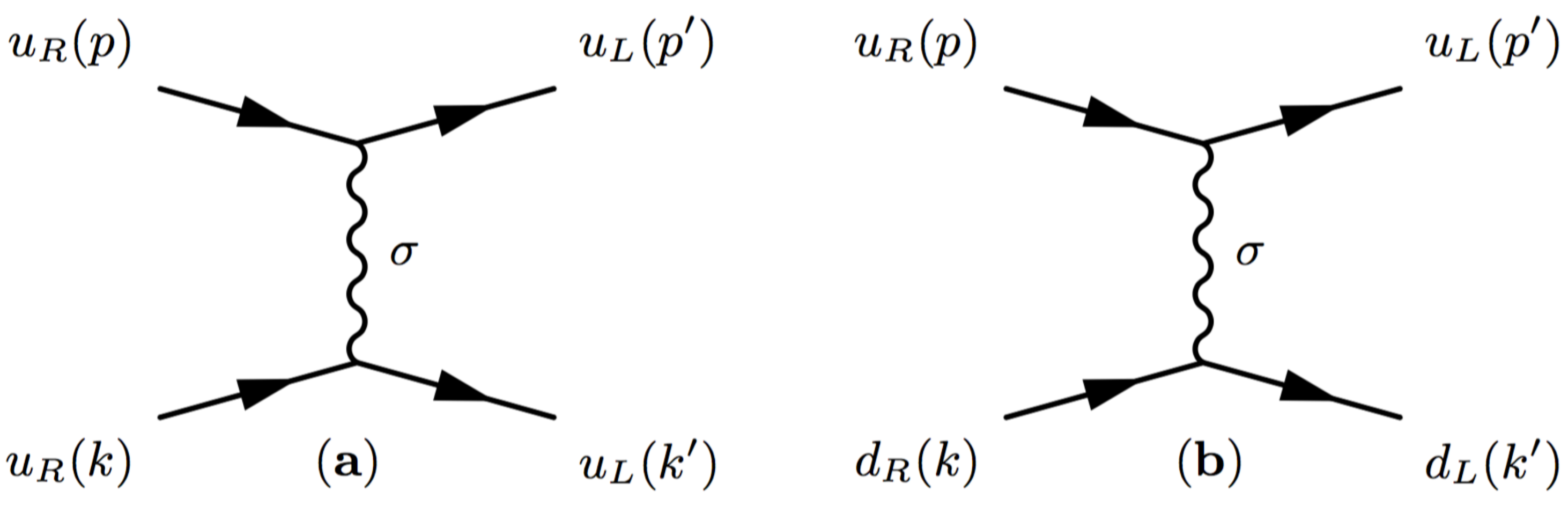}
\end{center}
\caption{\label{Fig:diagra2}Tree level diagrams for the chirality flips of $u$ quarks due to $\sigma$-exchange.}
\end{figure*}

For the scattering amplitude of chirality change due to $\sigma$-exchange we can follow the same lines of the
previous section. In this case the relavant diagrams are depicted in Fig.\ref{Fig:diagra2}, which we sum incoherently
in color and flavor. In the isospin symmetric limit however the two diagrams coincide. Instead of
Eq.\er{eq:msqEEE} we have
\begin{eqnarray}
\left|{\cal M}\right|_{\sigma qq}^2 = \frac{1}{4}\left[
4N_c(1+1)D_\sigma(t)D_\sigma^\dagger(t) a_1
+4N_c(1+1)D_\sigma(u)D_\sigma^\dagger(u)a_2 -2 N_c (1+1)D_{tu}
\left(a_1 + a_2- a_3\right)
\right],
\label{eq:msqEEEs}
\end{eqnarray}
with $D_\sigma$ given by Eq.\er{eq:mppSS},
$D_{tu} =D_\sigma(t)D_\sigma^\dagger(u) + D_\sigma(u)D_\sigma^\dagger(t)$ and $a_i$ defined as
in Eq.\er{eq:msqEEE} . For the real and imaginary part of $\Pi_\sigma$ the arguments given above for the
pion self-energy are still valid, hence we do not repeat them here. 

Before going ahead we remark that a full calculation would amount to consider the interference between the diagrams
for $\sigma$ and pion exchange. We do not do this in our work for simplicity; this decision is partly
justified {\it a posteriori} by the fact that we find the collision rate due to $\sigma$ exchange is always smaller
than the one due to pion exchange, hence we expect that the interference of the two processes does not affect
 considerably our results.

\section{Results}

\subsection{The NJL model}

\begin{figure}[t!]
\begin{center}
\includegraphics[width=8cm]{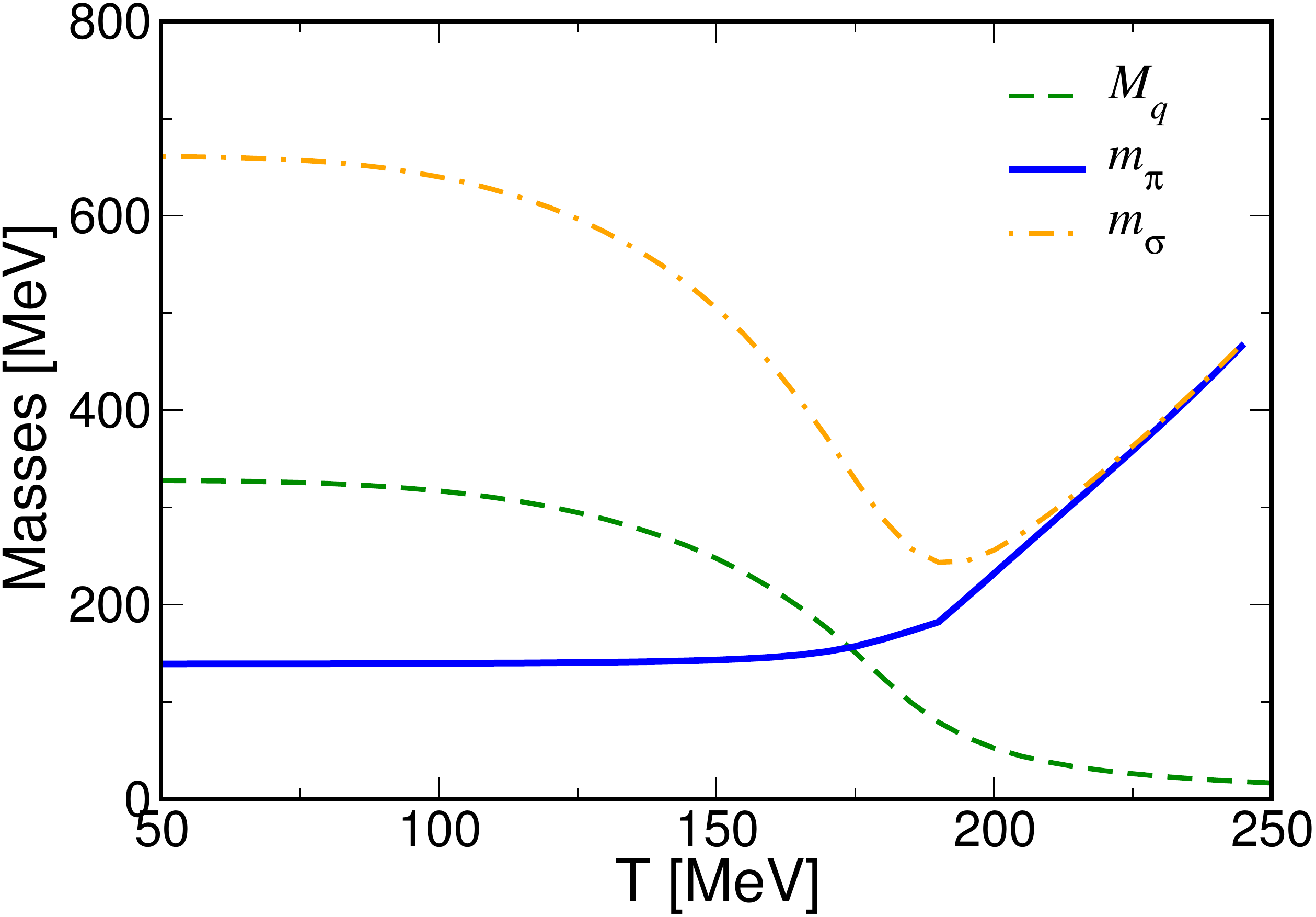}
\end{center}
\caption{\label{Fig:4a}$M_q$ (green dashed line), $m_\pi$ (solid blue line) and
$m_\sigma$ (orange dot-dashed line) versus temperature.}
\end{figure}

In Fig.~\ref{Fig:4a} we show by green dashed line $M_q$ versus temperature, 
computed by minimization of the thermodynamic potential
\sugg{in the NJL model}
at $\mu_5=0$ given by Eq.\er{eq:NJL3}. We need this quantity as it enters into the collision integral\er{eq:aplp}
via quark distribution functions and squared matrix element\er{eq:msqEEE}. It also enters into the
relation between $n_5$ and $\mu_5$ Eq.\er{eq:n5tyu}. From data shown in  Fig.~\ref{Fig:4a} we notice
a rapid decrease of $M_q$ in the temperature range $(150,200)$ MeV, connecting a low temperature phase
where chiral symmetry is spontaneously broken to a high temperature phase where chiral symmetry is approximately 
restored. In Fig.~\ref{Fig:4a}  we also plot our results for masses of pions and $\sigma$-meson, 
denoted respectively by $m_\pi$ and $m_\sigma$, for later reference.  By the inflection point of $M_q$
we can define a pseudo-critical temperature, $T_c\simeq 175$ MeV, for chiral symmetry restoration.

\begin{figure}[t!]
\begin{center}
\includegraphics[width=8cm]{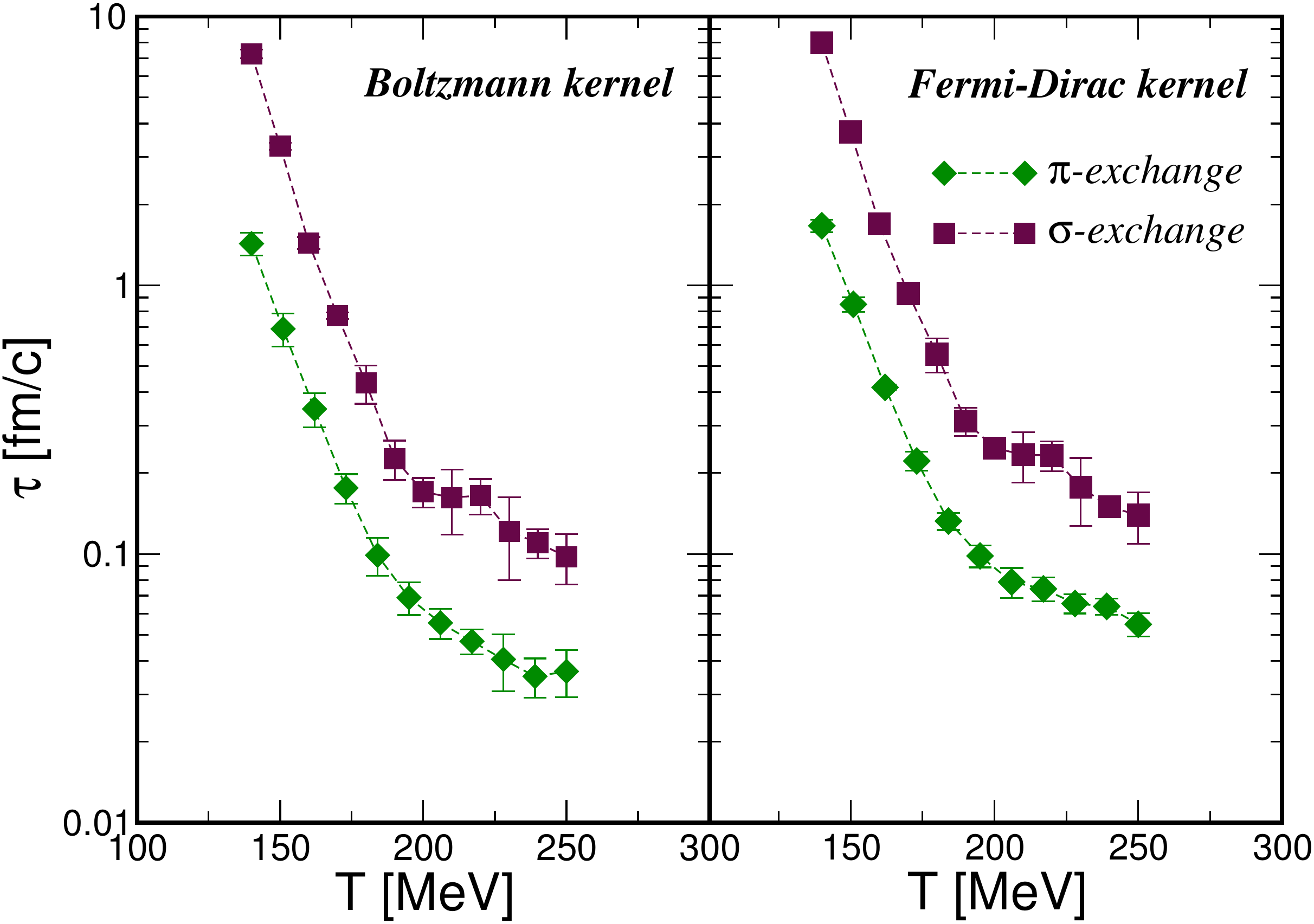}
\end{center}
\caption{\label{Fig:4}Relaxation time of chiral density for one-pion exchange (green diamonds) 
and $\sigma$-exchange (maroon squares) versus temperature. In the left panel we plot the results
obtained by the Boltzmann kernel; on the right panel we show our results obtained by
the Fermi-Dirac kernel.}
\end{figure}

Next we turn to the computation of the relaxation time of chiral density.
The main task is to compute the 12-dimensional integral in Eq.\er{eq:GAMMAequ}.
We use the 3-dimensional Dirac delta to perform the integral over $d^3 p^\prime$ trivially; 
a change of variables (namely a rigid rotation) allows to take $\bm p$ along the $z-$axis implying
$\int d^3 p = 4\pi \int p^2 dp$. 
The Dirac delta expressing energy conservation is used to integrate over $k_z$.
Eventually we are left with a 6-dimensional integral over $p^2 dp d^2 k_T d^3 k^\prime $
with $d^2 k_T = dk_x dk_y$. We perform this integral numerically by a quasi-Monte Carlo 
routine \cite{NIEDER,Caflisch} which uses 
the MISER Monte Carlo adaptive algorithm \cite{MISER,NR} with a Sobol low discrepancy 
sequence \cite{SOBOL} in place of a uniform random sequence. The integration variables are scaled
in units of temperature $T$, and we cutoff integrals at $10T$.

In Fig.~\ref{Fig:4} we plot the relaxation time of chiral density for one-pion exchange (the green diamonds) 
and $\sigma$-exchange (the maroon squares) versus temperature.  On the left panel we plot the results
obtained by the Boltzmann kernel in Eq.\er{eq:op1}; on the right panel we show our results obtained by
the Fermi-Dirac kernel in Eq.\er{eq:op2}. 
One of the most interesting aspects of data shown in Fig.~\ref{Fig:4} is the qualitative behavior of the relaxation
time versus temperature: we find that regardless the interaction channel chosen, as well as the statistics used in 
the collision integral, $\tau$ decreases with temperature. 
The lowering of $\tau$ is more evident in the low temperature phase and in the crossover region,
staying almost constant in the high temperature phase. 

In Fig.~\ref{Fig:4} we have shown the relaxation time for $\sigma$ mesons and for pions. We find that
the relaxation time of pions is always smaller than the one of $\sigma$ mesons: 
in the low temperature phase this is mainly due to the
larger mass of the latter in comparison with that of pions, see Fig.~\ref{Fig:4a}. In the high temperature phase,
where $m_\sigma \simeq m_\pi$, the relaxation time due to $\sigma$ exchange is still larger than the 
one obtained by pions: as a matter of fact, even if diagram ${\mathbf{(a)}}$ in Fig.\ref{Fig:diagra} is 
equal to diagram ${\mathbf{(a)}}$ in Fig.\ref{Fig:diagra2}, the multiplicity of diagrams for $\pi$-mediated
scattering is larger than the one for $\sigma$-exchange, implying the former has a larger scattering rate
and a smaller relaxation time.

It is interesting to compare the results obtained by using the Boltzmann kernel\er{eq:op1} in Eq.\er{eq:GAMMAequ}
with those obtained by the Fermi-Dirac kernel\er{eq:op2}, shown in Fig.~\ref{Fig:4} 
on left and right panel respectively. As expected, the use of the correct Fermi-Dirac
statistics leads to a slight increase of the relaxation time, corresponding to a lowering of the collision rate.
This is due to the Pauli blocking factors in the collision integral which effectively reduce the phase space available
for the collisions.

The response of $\tau$ to temperature might sound counterintuitive,
as one might expect that increasing temperature chiral symmetry gets restored so the processes able to flip
chirality of quarks, which are naively expected to be governed by $M_q$, should be suppressed and $\tau$ should increase.
However, a closer analysis of the problem shows that it is not so trivial and one has to consider carefully all the factors
governing the relaxation time, which are the phase space available for collisions on the one hand, and the 
interaction strength on the other hand. As a matter of fact, although the effective coupling strength 
among quarks and pions can decrease with temperature,
the phase space available for collisions becomes larger thanks to smaller quark masses 
and larger temperature which broadens the distribution functions.

\begin{figure}[t!]
\begin{center}
\includegraphics[width=8cm]{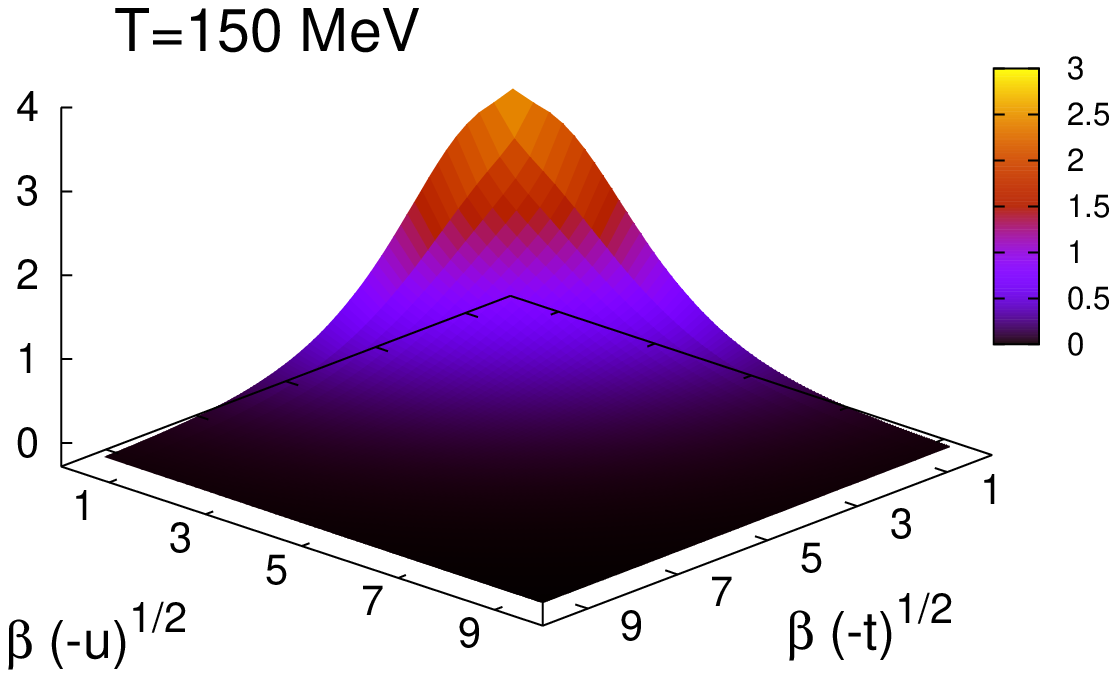}~~~~~~~~~~~~\includegraphics[width=8cm]{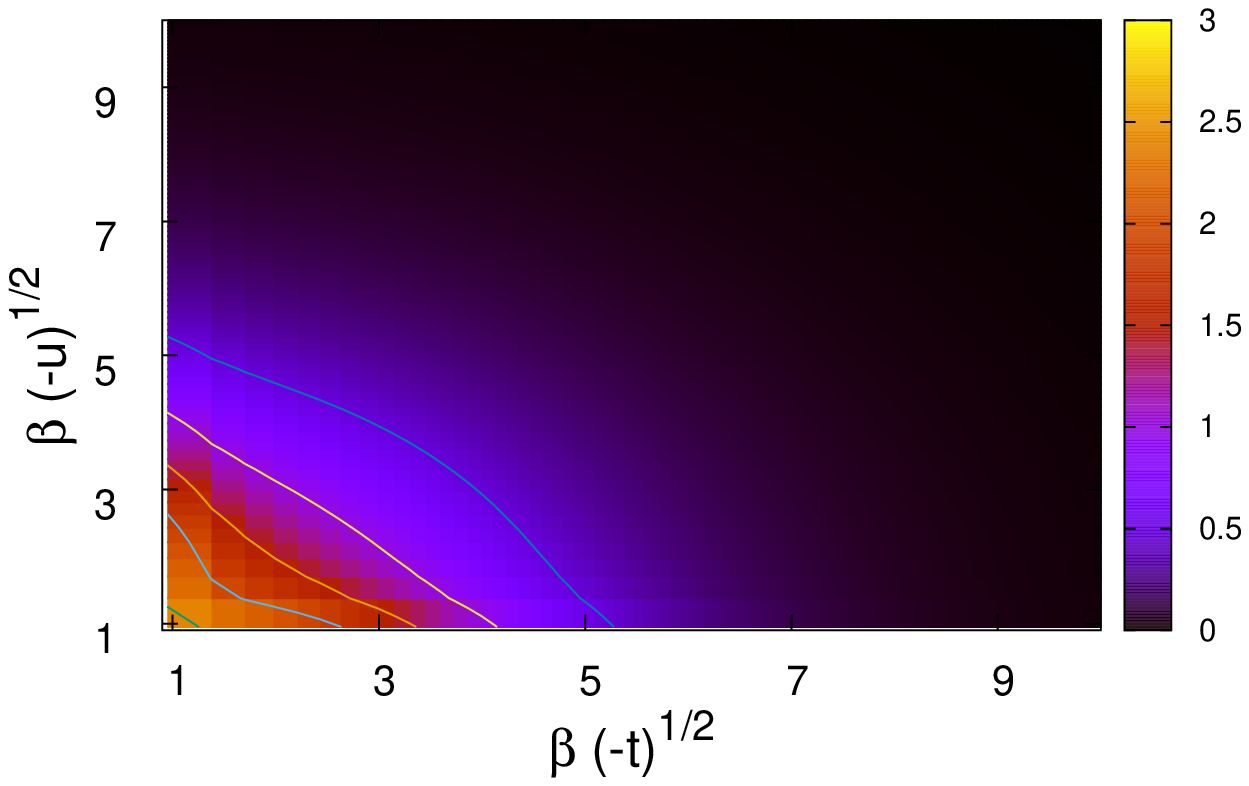}\\
\includegraphics[width=8cm]{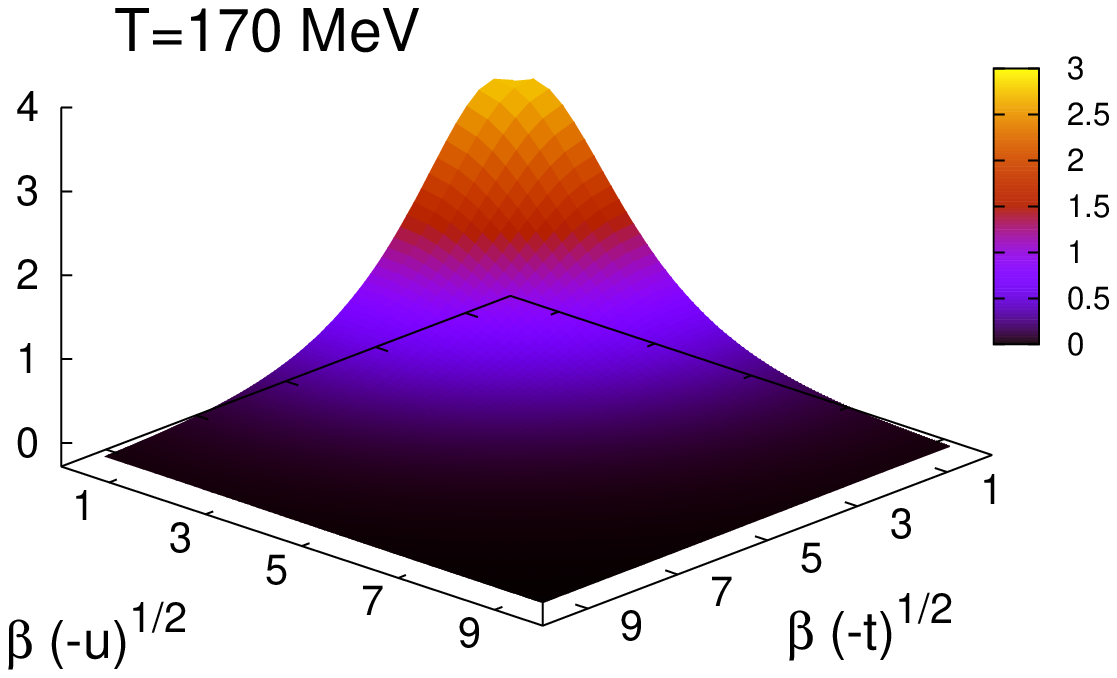}~~~~~~~~~~~~\includegraphics[width=8cm]{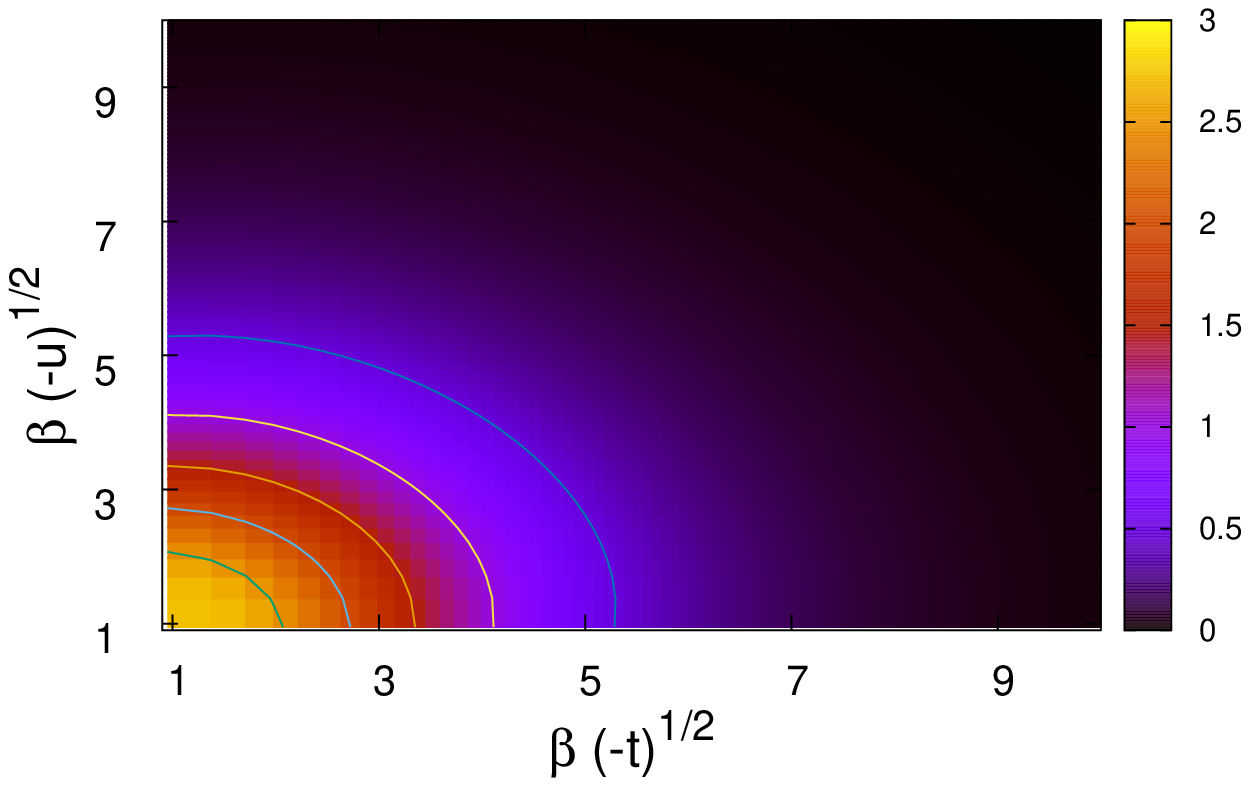}\\
\includegraphics[width=8cm]{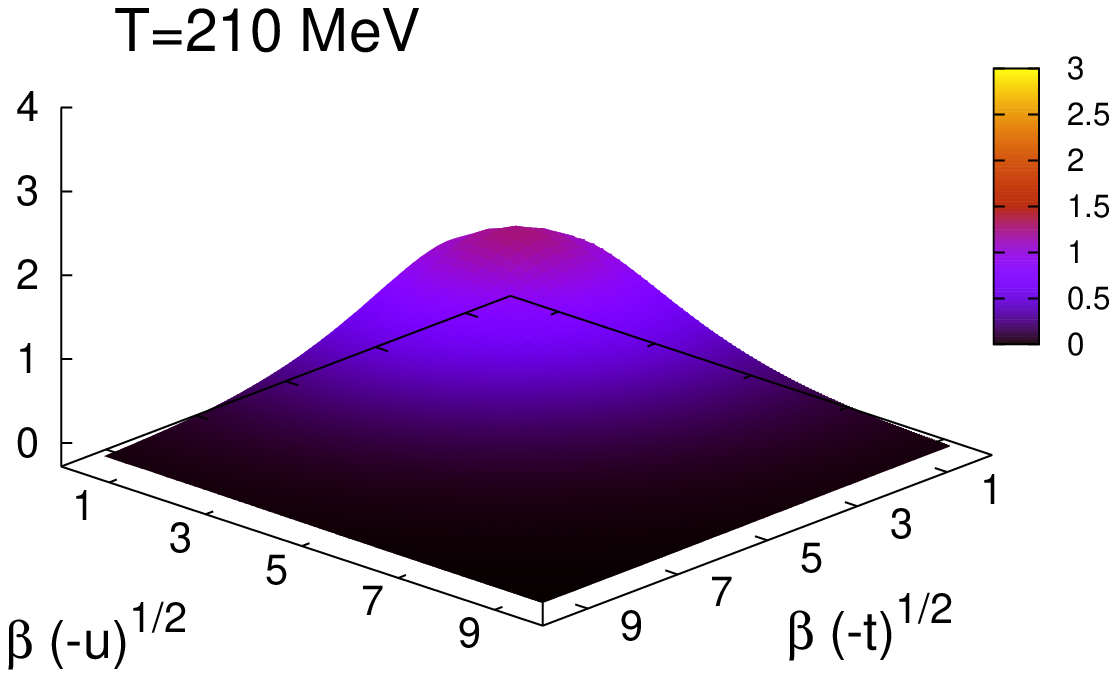}~~~~~~~~~~~~\includegraphics[width=8cm]{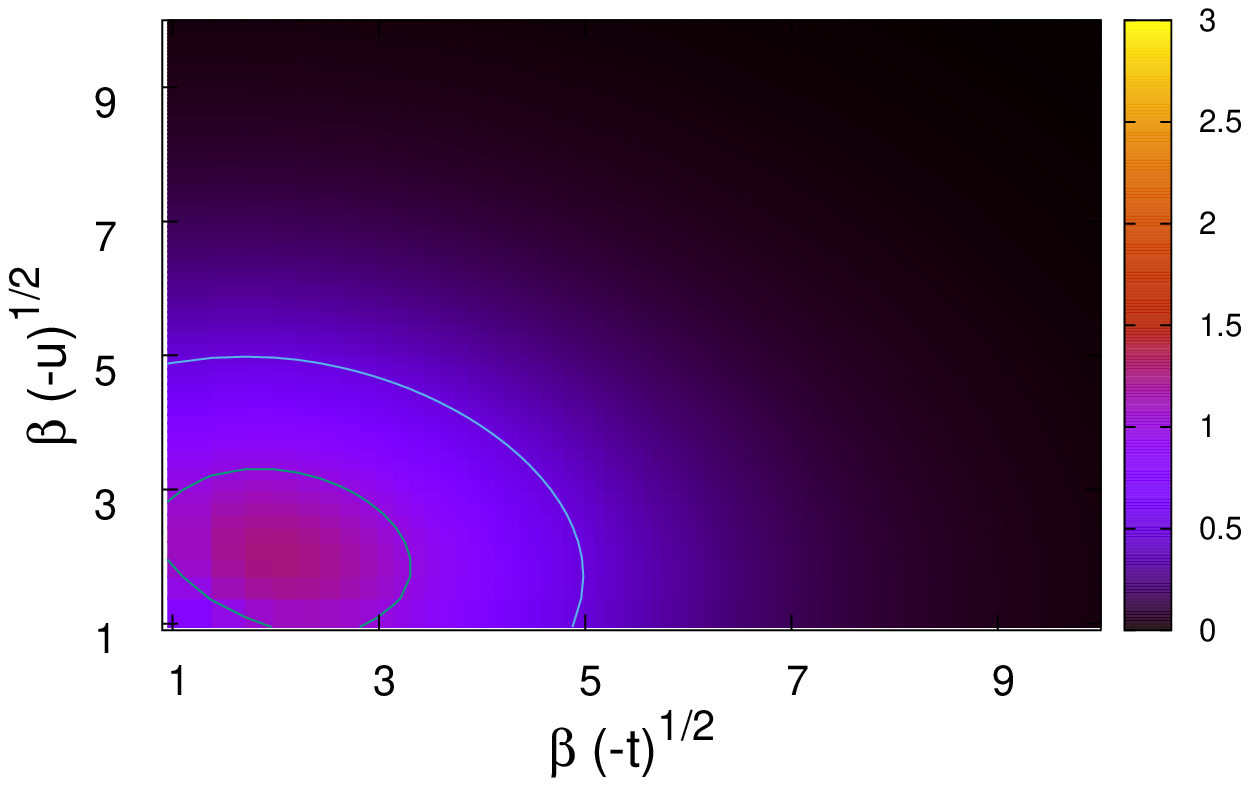}
\end{center}
\caption{\label{Fig:3d} \sug{The weighted matrix element} $J$, Eq.\er{eq:lpo} as a function of  
$\beta\sqrt{- t}$ and $\beta\sqrt{- u}$, with $\beta=1/T$, for several values of temperatures,
for the case of one pion exchange and Boltzmann kernel in the collision integral. \sug{The plots at the right column 
are contour representations of the corresponding quantities at the left column.}}
\end{figure}

\sug{This property} can be illustrated by computing 
\begin{equation}
J = |{\cal M}(t,u)|^2 \left.\frac{\partial F(t,u)}{\partial\mu_5}\right|_{\mu_5=0},
\label{eq:lpo}
\end{equation} 
where $(t,u)$ correspond to the Mandelstam variables and $F$ corresponds to the kernel of the 
collision integral either in Eqs.\er{eq:op1} or\er{eq:op2}, whose derivative with respect to $\mu_5$
at $\mu_5=0$ enters in the linearized collision rate\er{eq:GAMMAghj}. 
\sug{The quantity} $J$
can be interpreted as the squared matrix element weighted by distribution functions.
We limit this discussion to the case of $\pi$-mediated scattering and to the Fermi-Dirac
kernel in Eq.\er{eq:op1}, since other cases do not differ qualitatively from this one. 
In Fig.~\ref{Fig:3d} we plot
$J$ versus $\beta\sqrt{- t}$ and $\beta\sqrt{- u}$ with $\beta=1/T$ for several values of temperatures,
for the case of one pion exchange and Fermi-Dirac kernel in the collision integral. Plots on the right column 
are contour representations of the same quantities shown on the left column. 
From upper to lower panels we plot $J$ for $T=150$ MeV which is below $T_c$, for $T = T_c$, 
and finally, for $T=210$ MeV.

We notice that increasing temperature the magnitude of \sug{weighted matrix element} $J$ becomes 
gradually smaller, hence the
scattering matrix itself becomes less efficient in producing chirality changes in the thermal bath.
On the other hand, $J$ spreads in momentum space as temperature is increased: 
in fact from the data shown in the contour plots in the
figure results that increasing temperature  $J$ gets its larger contribution from the square 
$0\leq \beta\sqrt{-t} \leq 5$, $0\leq \beta\sqrt{-u} \leq 5$ in momentum space, which covers a fraction
of phase space growing as $T^2$ with temperature.
As a consequence, the amount of phase space occupied
by quarks and giving a contribution to the collision integral increases with temperature, competing against
the lowering of the scattering matrix and eventually leading
to the increase of the collision rate.

\begin{figure}[t!]
\begin{center}
\includegraphics[width=8cm]{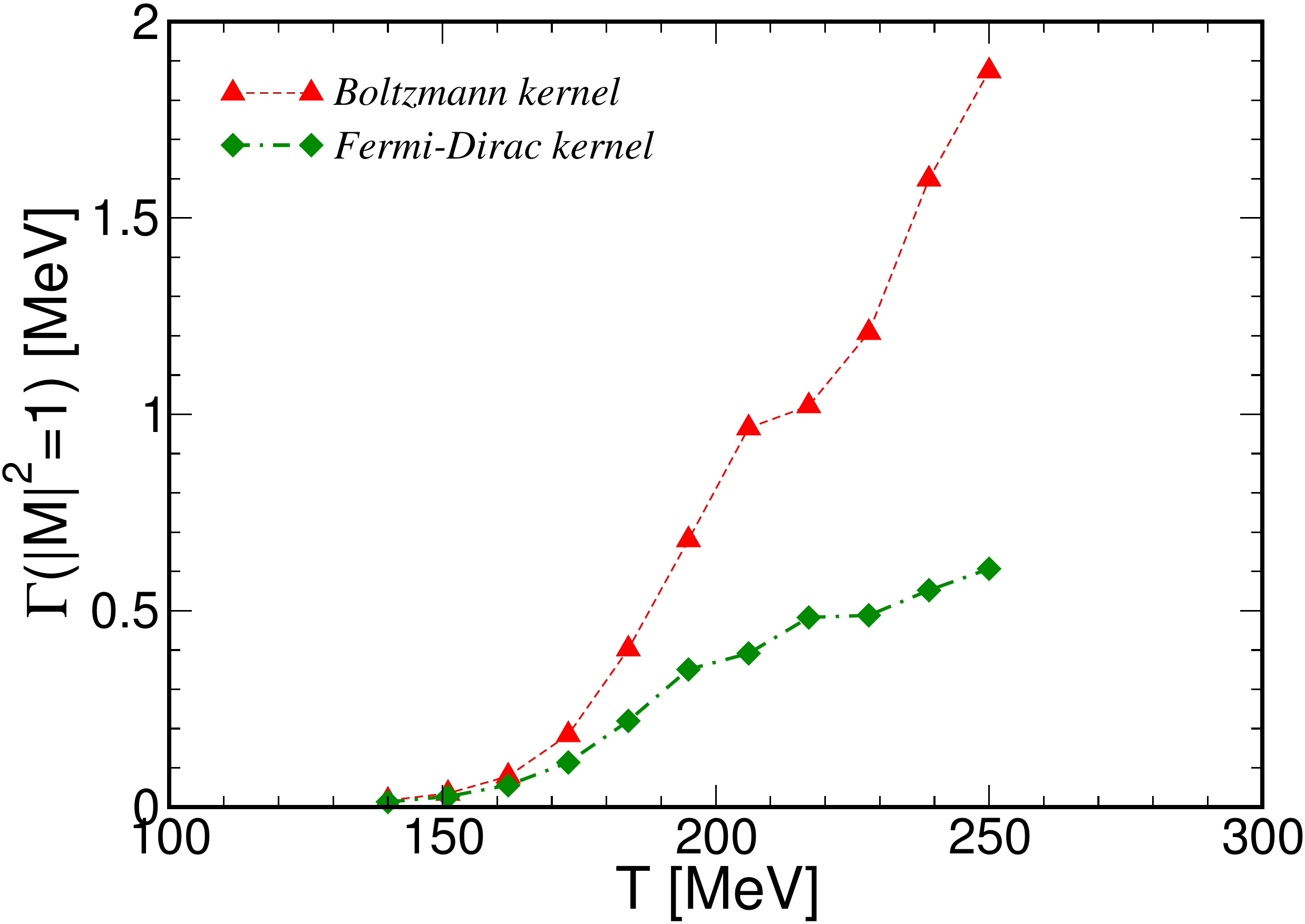}\\
\includegraphics[width=8cm]{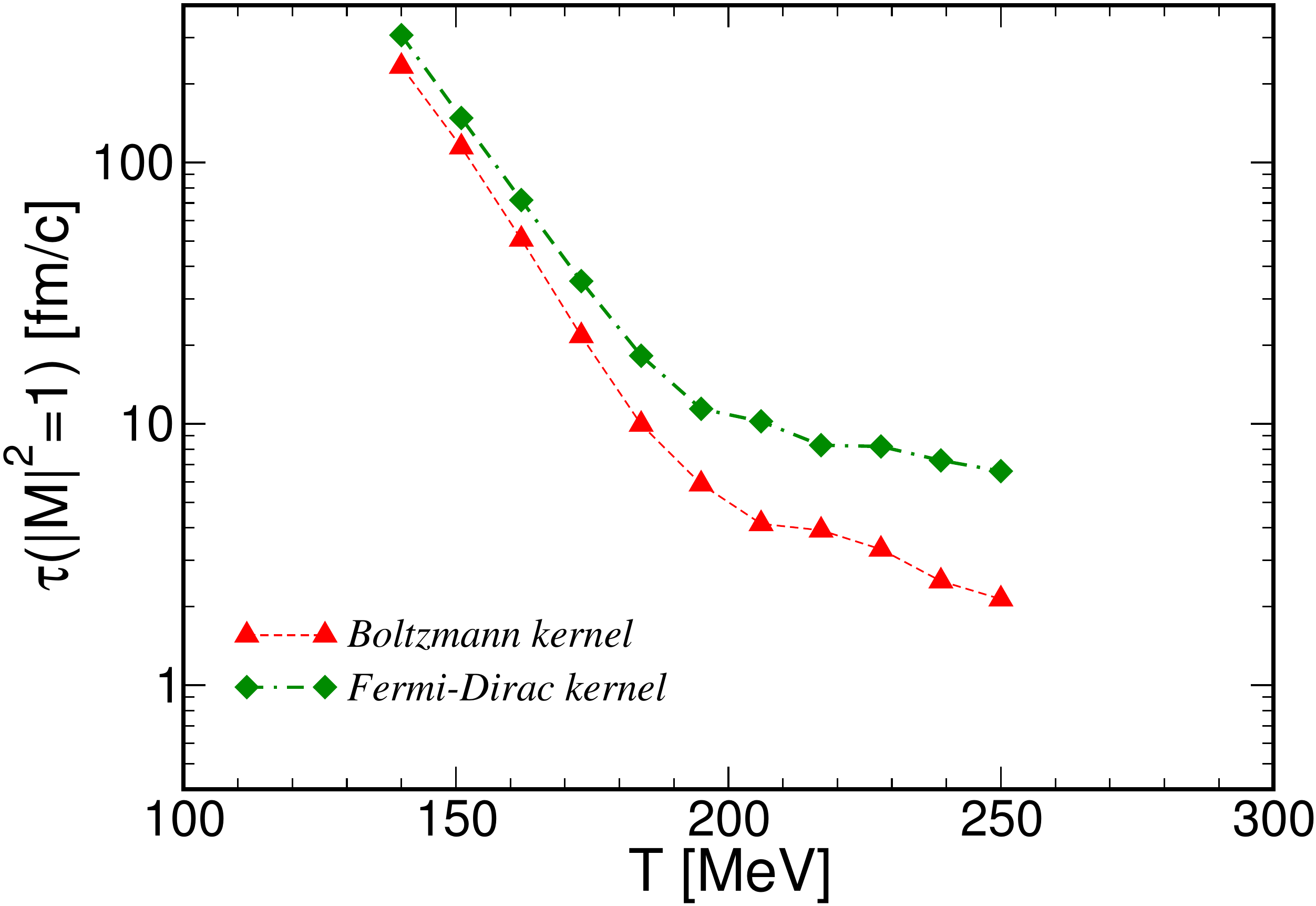}
\end{center}
\caption{\label{Fig:4d}Upper panel: Collision rate with $|{\cal M}|^2=1$ for Boltzmann (red triangles) and
Fermi-Dirac (green squares) kernels. Lower panel: relaxation time with $|{\cal M}|^2=1$.}
\end{figure}
We can elaborate more on the role of the effective phase space opening with temperature by computing the
collision rate with $|{\cal M}|^2 = 1$ in Eq.\er{eq:PS2}: in this way we remove every detail about collisions,
and $\Gamma$ reacts only to the variation of the distribution functions hence measuring the amount of
momentum space involved in the collisions. In Fig.~\ref{Fig:4d} we plot  $\Gamma$ with $|{\cal M}|^2 = 1$
versus temperature for the cases of Boltzmann (red triangles) and Fermi-Dirac (green squares) statistics.
In both cases we find a noticeable increase of this $\Gamma$ in the crossover region: changing $T$
from 150 to 200 MeV we find $\Gamma(T=200)/\Gamma(T=150) \simeq 18$ for the case of the Boltzmann
kernel, and $\Gamma(T=200)/\Gamma(T=150) \simeq 11$ for the case of the Fermi-Dirac kernel.

\subsection{Hybridization of NJL with a quasiparticle model}
Although the NJL model offers a nice qualitative description of the chiral crossover at finite temperature, 
it is likely to miss the description of relevant degrees of freedom above $T_c$. As a matter of fact,
quarks in the NJL model above $T_c$ have in the chiral limit a vanishing mass within the mean field approximation;
on the other hand it is known that at large temperatures quarks develop a chirally invariant self-energy
which generates a thermal pole mass, $M_T\propto g T$, due to the QCD 
interactions, and this thermal mass increases with temperature. Evading the chiral limit {\sug {by}} adding a small current quark mass
in the NJL model does not change the fact that $M_q$ decreases with temperature and $M_q\ll T$ for $T>T_c$.
There are however quasiparticle models  inspired by the behavior of the thermal mass in QCD at $T\gg T_c$,
in which one assumes that a quasiparticle description is valid also at temperatures
$T\simeq T_c$, see for example~\cite{Gorenstein:1995vm,Plumari:2011mk,Levai:1997yx,Bluhm:2004xn,Bluhm:2007nu,Castorina:2011ja,Alba:2014lda,Ruggieri:2012ny,Oliva:2013zda,Begun:2016lgx}:  
in these models typically one assumes $M_T\propto gT$ with $g$ corresponding to 
\sug{a temperature dependent strong coupling constant 
fixed by a numerical fit to lattice data  from $T\simeq T_c$
up to very large temperatures}.

\begin{figure}[t!]
\begin{center}
\includegraphics[width=8cm]{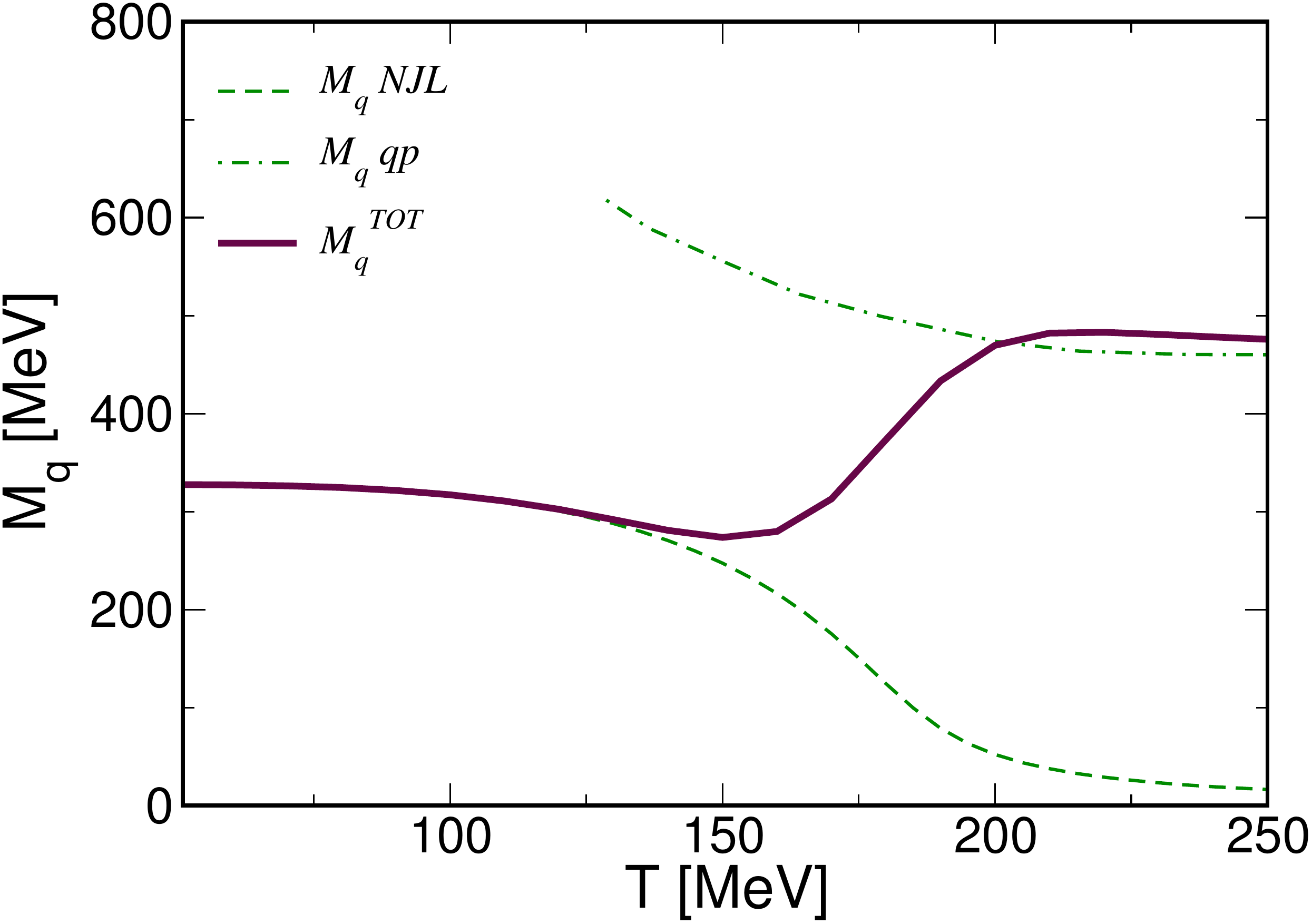}
\end{center}
\caption{\label{Fig:4hjkl}Constituent quark mass computed by the NJL model (the dashed line), quasiparticle 
thermal mass (dot-dashed line) and interpolation among the two (the solid line) used in the calculation of the
relaxation time.}
\end{figure}
In Fig.~\ref{Fig:4hjkl} we plot the constituent quark mass \sug{$M_q$} computed by the NJL model (the dashed line), compared to
the quasiparticle thermal mass \sug{$M_T$} (the dot-dashed line) computed in Ref.~\cite{Plumari:2011mk}.
\sug{The main difference between these two masses, $M_q$ and $M_T$} is that the latter \sug{-- contrary to the former --} is not related to a term $\propto \bar\psi\psi$ in the quark Lagrangian, being rather related to many body effects at finite temperature which induce a pole in the full quark propagator. While this concept is rigorous at very large temperature, in quasiparticle models one assumes for simplicity that the
pole mass is still a meaningful concept at the chiral crossover.  

Assuming the point of view of a quasiparticle model implies that above 
$T_c$ quark mass can be quite large even if chiral symmetry is restored: this can affect the collision rate
because of the reduction of momentum space occupied by quarks. We find therefore interesting
to compute the relaxation time of chiral density assuming a quasiparticle nature of quarks above $T_c$.
We achieve this by using a model for the quark mass which interpolates 
between the low-temperature NJL quark mass $M = M_q(T)$ and the high-temperature thermal quark mass 
$M_T = M_T(T)$,
shown by the solid orange line in Fig.~\ref{Fig:4hjkl}. The interpolating function we use is
\begin{equation}
\sug{M(T) = M_q(T) + a(T)M_T (T)},
\label{eq:ModelMass1}
\end{equation}  
where the function $a(T)$ is given by
\begin{equation}
a(T) =\frac{1}{2}\left[1 + \tanh\left(\frac{T-T_0}{c}\right)\right],
\label{eq:interpo}
\end{equation}
In the above equation $M_q$ corresponds to the solution of the NJL gap equation, and $M_T$ is the
quasiparticle mass obtained by fit of the \sug{corresponding lattice data} in~\cite{Plumari:2011mk}.
The two numerical parameters are chosen as $T_0 = 180$ MeV and $c=20$ MeV.
The functional form in Eq.\er{eq:ModelMass1}  is not the solution of a gap equation: 
it is chosen only to interpolate smoothly between 
$M_q$ and $M_T$ in the crossover region, with the purpose to illustrate the effect of
turning from the NJL model to the quasiparticle one on the relaxation time.

\begin{figure}[t!]
\begin{center}
\includegraphics[width=8cm]{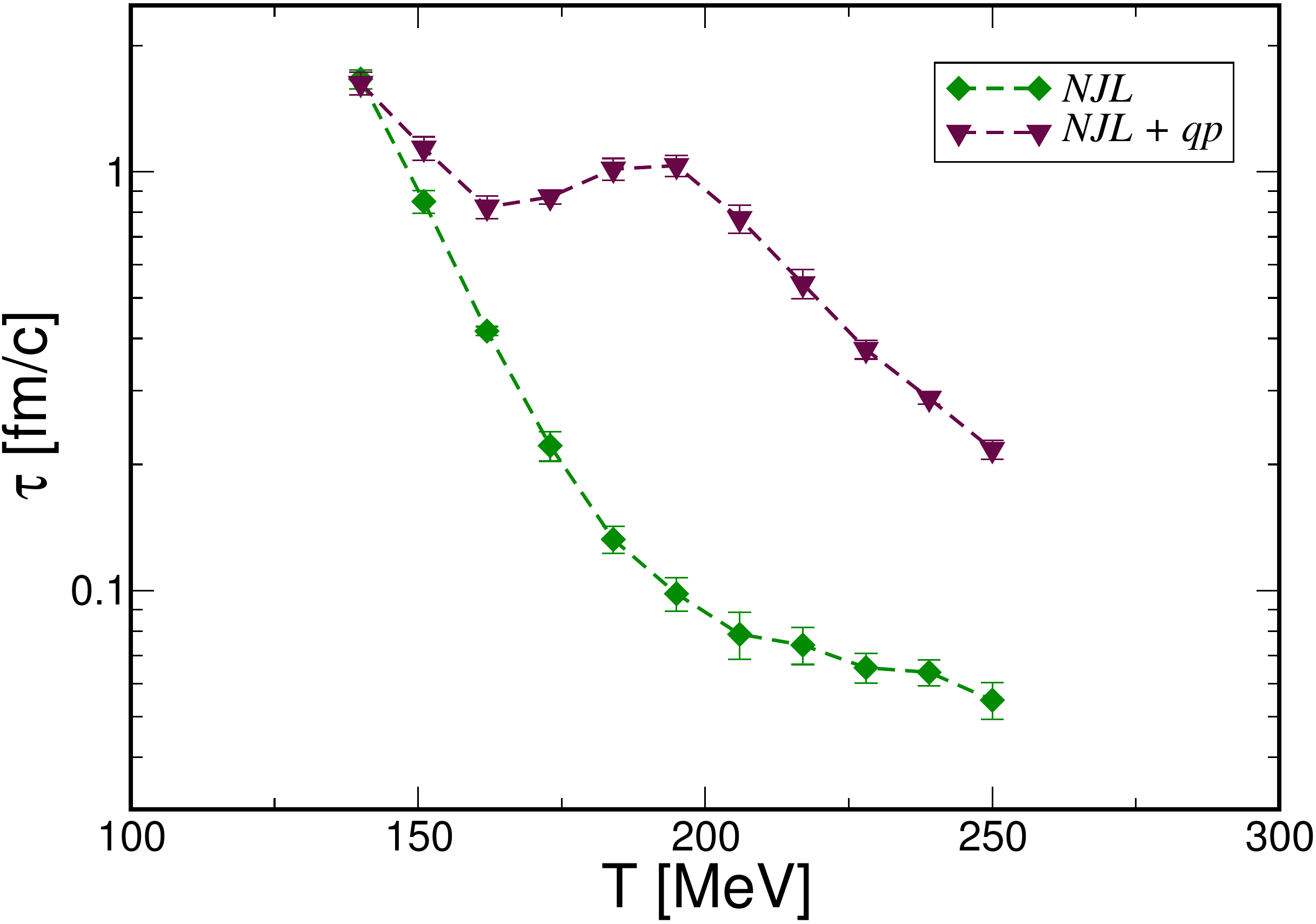}
\end{center}
\caption{\label{Fig:4hjklm}
\sug{The chiral relaxation time for the NJL model (the green diamonds) and
NJL{+}qp model (the maroon triangles).}}
\end{figure}

In Fig.~\ref{Fig:4hjklm} we plot the relaxation time versus temperature
for the cases of the pure NJL model (green diamonds) and
NJL hybridized with a quasiparticle model (maroon triangles); in both cases we have used the Fermi-Dirac
kernel in the collision integral. We find that the overall effect of the quasiparticle model mass is to increase the
relaxation time: as a matter of fact the increase of quark mass induces a lowering of the available phase space
for the collisions, leading to a smaller collision rate and a larger relaxation time in comparison with
the NJL calculation.
Although relaxation time within the quasiparticle model is larger than the one we obtain within the NJL model,
the effect of increasing temperature leads to a non monotonic behavior of the \sug{relaxation rate} $\tau$: in fact the increasing of $M$
given by Eq.\er{eq:ModelMass1} has to compete with the increase of temperature that opens phase space
and increases the collision rate.

\section{Conclusions}
In this article we have studied relaxation of chiral density, $n_5$, in 
the two flavor
Nambu-Jona-Lasinio (NJL) model
which describes in simple terms the chiral crossover of QCD
at a temperature $T_c\simeq 170$ MeV. In particular, we have computed the relaxation time for chiral density $\tau$ (``the chiral relaxation time'')
which is associated with the 
\sugg{scattering of quarks via one-pion and one-$\sigma$-meson exchanges.}
\sugg{These processes are good candidates for inducing chirality flips in quark matter around the chiral crossover within the effective-model approach based on the NJL model.}

In order to compute the 
\sugg{chiral} relaxation time, $\tau$, 
\lt{we have followed the well established formalism to deal with quark scattering within the NJL model. 
Firstly, we evaluated the finite-temperature $\pi$- and $\sigma$-meson propagators within the random phase approximation.
Secondly, we calculated scattering amplitudes due to meson exchanges. Thirdly, we used the latter to compute the collision integral for the chirality changing processes which is directly related to the relaxation of the chiral density, $d n_5/d t$, and we used again the NJL model to relate the chiral density $n_5$ with chiral chemical potential $\mu_5$ in the thermal bath. We assumed a weak chiral imbalance $\mu_5 \ll T$ in order to be able to work at the lowest order in $\mu_5/T$ what allowed us to drastically simplify the computation of the collision integral. Finally, we compute the collision rate $\Gamma$ and the chiral relaxation time $\tau$ via the relation $\Gamma = - (d n_5/d t)/n_5$. We focused on a temperature range around the chiral crossover because at this region the quark degrees of freedom should have solid physical meaning.}

\sugg{We found that the results for the chiral relaxation time $\tau$ do not depend on the statistics used to calculate the collision integral as, according to Fig.~4, both Boltzmann and Fermi-Dirac distributions give very similar results  in the chiral crossover region. 
Moreover, the same Figure demonstrates that the chiral relaxation time $\tau_\sigma$ due 
to the $\sigma$-meson exchange is much larger compared to the relaxation time $\tau_\pi$ 
corresponding to the pion exchanges, $\tau_\sigma \gg \tau_\pi$. This feature can be explained by the fact that in the low temperature phase the $\sigma$ exchange is suppressed because $\sigma$-meson mass is larger than pion mass. Around and above the chiral crossover the masses are of the same order, $m_\sigma\simeq m_\pi$, but still the relaxation time related to the $\sigma$-exchange is much larger compared to the one of the pion exchange due to larger number of the diagrams that contribute to the latter. Thus, the pion exchanges are dominating the chiral relaxation processes.}

\sugg{We have computed also the relaxation time in hybridized NJL model. 
where the constituent quark's mass in the chirally restored region is tuned to the thermal mass of the quarks obtained 
by a fit to lattice data about QCD thermodynamics in \cite{Plumari:2011mk}. 
Our results for the chiral relaxation time $\tau$ in the chiral crossover region are summarized in Fig.~\ref{Fig:4hjklm}. 
We find that regardless of the choice of the thermal quark mass, the chiral relaxation time follows an almost monotonic behavior with increasing temperature, even if the effect of the thermal mass
is to keep $\tau$ higher compared to the one computed within the NJL model.
Globally, the relaxation time falls down with increase of the temperature from $\tau \simeq 1$\,fm at the lower-temperature end of the crossover at $T \simeq 150$\,MeV and till much faster chiral flips, $\tau \simeq 0.1$\,fm at the higher-temperature at $T \simeq 250$\,MeV.}
\sugg{The fast increase of the collision rate (i.e. the lowering of $\tau$) 
with rising temperature can be understood as a combination of two factors: 
on the one hand,
 the scattering matrix weighted by the distribution functions decreases with temperature, but on the other hand it also
broadens in the momentum space thus effectively leading to a growing of the phase space volume involved in
collisions. 
The latter dominates over the former, thus enhancing the collision rate and lowering the relaxation time 
with increase of temperature.}


\begin{acknowledgments}
M. R and G. X. P. would like to thank the 
CAS President's International Fellowship Initiative (Grant No. 2015PM008), 
and the NSFC projects (11135011 and 11575190). M. R. acknowledges discussions
with F. Scardina. Source code for the implementation
of pseudo-random and low discrepancy sequences used in our numerical calculations
are distributed under the GNU-LGPL license and can be found on the John Burkardt
\href{http://people.sc.fsu.edu/~jburkardt/f_src/f_src.html}{Fortran 90 Source Codes}
website.
\end{acknowledgments}

\appendix

\end{document}